\documentclass[traditabstract]{aa} 
\usepackage{graphicx}
\usepackage{psfrag}
\usepackage{amsmath}
\usepackage{amssymb}
\usepackage{color,xcolor}
\usepackage{hyperref}
\usepackage{caption,natbib}
\hypersetup{
    colorlinks=true,
    citecolor=blue,
    filecolor=black,
    linkcolor=blue,
    urlcolor=magenta,
    linktocpage=true,
    breaklinks = true
}

\usepackage{enumitem}
\usepackage{subfigure,float}
\usepackage{multirow}
\usepackage{makecell}
\usepackage[normalem]{ulem}

\usepackage{array,booktabs}
\usepackage{siunitx}
\DeclareUnicodeCharacter{2212}{\textendash}

\newcommand{\beq}{\begin{equation}}
\newcommand{\eeq}{\end{equation}}

\newcommand{\ber}{\begin{eqnarray}}
\newcommand{\eer}{\end{eqnarray}}

\newcommand{\ba}{\begin{align}}
\newcommand{\ea}{\end{align}}

\usepackage{orcidlink} 
\usepackage{url}
\usepackage{placeins}

\usepackage{tabularx}  
\usepackage{placeins}
\begin{document}

   \title{Detection of Unresolved Strongly Lensed Supernovae with 7-Dimensional Telescope}

   \author{Elahe Khalouei \inst{1}\orcidlink{0000-0001-5098-4165}\thanks{e.khalouei1991@gmail.com}
   , Arman Shafieloo\inst{2,3}\orcidlink{0000-0001-6815-0337}\thanks{shafieloo@kasi.re.kr} , Alex G. Kim\inst{4}\orcidlink{0000-0001-6315-8743}, Ryan E. Keeley\inst{5}\orcidlink{0000-0002-0862-8789},  William Sheu\inst{6}\orcidlink{0000-0003-1889-0227},  Gregory S. H. Paek\inst{1,7}\orcidlink{0000-0002-6639-6533}, Myungshin Im\inst{1,7}\orcidlink{0000-0002-8537-6714},  Xiaosheng Huang\inst{8,9}\orcidlink{0000-0001-8156-0330}, Hyung Mok Lee\inst{1}\orcidlink{0000-0003-4412-7161}}
 
   \institute{Astronomy Research Center, Research Institute of Basic Sciences,
Seoul National University, 1 Gwanak-ro, Gwanak-gu, Seoul 08826,
Korea
         \and 
         Korea Astronomy and Space Science Institute (KASI), 776 Daedeok-daero, Yuseong-gu, Daejeon 34055, Korea
           \and
           KASI Campus, University of Science and Technology, 217 Gajeong-ro, Yuseong-gu, Daejeon 34113, Korea
           \and
           Lawrence Berkeley National Laboratory, 1 Cyclotron Road, Berkeley, CA 94720, USA
          \and
           Department of Physics, University of California Merced, 5200 North Lake Road, Merced, CA 95343, USA
          \and
          Department of Physics \& Astronomy, University of California, Los Angeles 430 Portola Plaza, Los Angeles, CA 90095, USA
          \and
          Astronomy Program, Department of Physics and Astronomy, SNU, Seoul, Republic of Korea
          \and
          Department of Physics \& Astronomy, University of San Francisco 2130 Fulton Street, San Francisco, CA 94117-1080, USA 
          \and
          Physics Division, Lawrence Berkeley National Laboratory 1 Cyclotron Road, Berkeley, CA 94720, USA
}

\abstract
{Gravitationally lensed supernovae (glSNe) are a powerful tool for exploring the realms of astronomy and cosmology. Time-delay measurements and lens modeling of glSNe can provide a robust and independent method for constraining the expansion rate of the universe. The study of unresolved glSNe light curves presents a unique opportunity for utilizing small telescopes to investigate these systems. In this work, we investigate diverse observational strategies for the initial detection of glSNe using the 7-Dimensional Telescope (7DT), a multitelescope system composed of twenty 50-cm telescopes. We implement different observing strategies on a subset of 5807 strong lensing systems and candidates identified within the Dark Energy Camera Legacy Survey (DECaLS), as reported in various publications. 
Our simulations under ideal observing conditions indicate the maximum expected annual detection rates for various glSNe types (Type Ia and core-collapse (CC)) using the 7DT target observing mode in the 
$r$-band at a depth of 22.04 mag, as follows: 
7.46 events for type Ia, 2.49 for type Ic, 0.8 for type IIb, 0.52 for type IIL, 0.78 for type IIn, 3.75 for type IIP, and 1.15 for type Ib. Furthermore, in the case of medium-band filter observations (m6000) at a depth of 20.61 in the
Wide-field Time-domain Survey (WTS)
program, the predicted detection rate for glSNe Ia is 2.53 $yr^{-1}$.
Given targeted follow-up observations of these initially detected systems with more powerful telescopes, we can  
apply a model-independent approach to forecast the ability to measure $H_{0}$ using a Gaussian process from Type Ia Supernovae (SNe Ia) data and time-delay distance information derived from glSNe systems, which include both Ia and CC types. 
We forecast that the expected detection rate of glSNe systems  
can achieve a $2.7\%$ precision in estimating the $H_{0}$.
}
   {}
   {}
   {}
   {}


   \titlerunning{Cosmography with 7DT}
   \authorrunning{Khalouei et al}

   \maketitle

\section{Introduction}
The Lambda Cold Dark Matter ($\Lambda$CDM) model serves as the standard framework in cosmology. This model provides explanations for a wide array of current observations, including the cosmic microwave background (CMB) radiation and baryon acoustic oscillations (BAO) \citep{Schlegel,Ade1,Ade2,aghanim,alam}.  
Despite its successes, the model struggles to resolve differences in measurements of the current expansion speed of universe \citep{DiValentino}.
These discrepancies arise when comparing estimates derived from observations of the early universe 
\citep{aghanim}
with those obtained from local observations, such as Type Ia supernovae (SNe Ia) calibrated by Cepheid variable stars
\citep{Riess}.

Gravitationally lensed transients, like quasars (QSOs) and  supernovae (SNe)
operate as independent cosmological probes capable of constraining the Hubble constant ($H_{0}$) (\citet{Treu}, and references therein). 
This is achieved through direct estimation of $H_{0}$ using time-delay measurements in combination with precise lens modeling
\citep{Refsdal64,Refsdal264,Oguri2006,Birrer12020,Birrer2020,kelly2,pascala}.
The abundance of lensed QSOs makes them a consistent and reliable resource for time-delay cosmography. Although hundreds of lensed QSOs have been identified \citep{lemon}, accurately measuring their time delays remains a challenging task \citep{Liao}. This difficulty arises from their stochastic light curves and variability on year-long timescales, requiring prolonged monitoring with high-resolution telescopes, which is both time-consuming and costly \citep{milon}. As a result, only a small fraction of lensed QSOs have been utilized for cosmological studies (e.g., \citealt{wong,shajib,Shajib2} ). 
Gravitationally lensed supernovae (glSNe) play a pivotal role in cosmology, offering several distinct advantages. Their well-defined light curves enable precise measurements of time delays and $H_0$.
In particular, the unique properties of SNe Ia as standard candles allow for direct measurement of intrinsic luminosities, provided that the magnification effects caused by microlensing from stars in the foreground lens galaxy can be mitigated \citep{Foxley,Weisenbach}.
Moreover, \citet{birrer} highlight that the standardizable brightness of  glSNe Ia provides tighter constraints on lens mass models, effectively addressing the mass-sheet degeneracy \citep{Falco,Schneider} and reducing systematic uncertainties in the determination of $H_0$.
Additionally, follow-up observations after the glSNe have faded enable detailed studies of stellar kinematics and host galaxies \citep{Ding, Suyu2023}.
\\
\indent To date, a total of nine SNe have been confirmed to be strongly gravitationally lensed. These include PS1$-$10afx \citep{Qu}, SN Refsdal \citep{kelly1}, SN 2016geu \citep{goobar}, SN Requiem \citep{Rodeny}, AT 2022riv \citep{ATr}, SN Zwicky \citep{goobar2,pier}, C22 \citep{chen}, SN $H0pe$ \citep{Frye, Poll}, and SN Encore \citep{Pierelen, Dhawan}.
Each of these SNe is lensed by either a single galaxy or a cluster of galaxies.  The SN Refsdal, observed at a redshift of 1.49, initially presented four lensed images in 2014, followed by the detection of a fifth image in 2015 \citep{kelly1}. Utilizing the time-delay data between these images, \citet{kelly3,kelly2} conducted an analysis that estimated $H_{0}$ to be $66.6_{-3.3}^{+4.1}$ km s$^{-1}$Mpc$^{-1}$.
This work demonstrates the use of gravitational lensing phenomena to enhance our understanding of cosmic scale parameters. 
Recently, the $SNH0pe$ is identified as the first gravitational lensing system discovered  with the James Webb Space Telescope (JWST). This system is magnified by the galaxy cluster $PLCK-G165.7+67.0$
\citep{Frye, Poll}. 
\citet{pascala} have presented the first measurement of $H_{0}$$= 75.7_{-5.5}^{+8.1}$ km s$^{-1}$Mpc$^{-1}$
from $SNH0pe$. 
\\
Simulation studies indicate that, given the limiting magnitude threshold and observing strategy, hundreds of glSNe could be discovered each year with 
the Rubin Legacy Survey of Space and Time (LSST; see, e.g., \citealt{LSSTrev} for a review).
\citet{pierel} forecast that the Roman Space Telescope will significantly advance the discovery of glSNe systems, encompassing both Type Ia and Core-Collapse (CC) SNe. 
Their study demonstrates that various types of glSNe can refine cosmological parameters, including the $H_{0}$, providing strong motivation for focusing observational studies on different types of glSNe events. 
The Roman Space Telescope, with an angular resolution of approximately 0.11 arcseconds, will be capable of resolving glSNe images with significantly smaller separations compared to LSST, which has an angular resolution of about 0.5 arcseconds. Furthermore, Roman will primarily detect glSNe at higher redshifts, making it especially valuable as a high-redshift survey instrument for cosmographic studies utilizing glSNe \citep{pierel}.
\\
\indent Recently, imaging surveys have identified thousands of new strong lenses and candidates, with the majority being galaxy-scale lenses, along with a smaller number of group or cluster lenses (e.g., \citealt{ Jacobs, H2020, H2022, can21, can22,shu221,stein,sheu,Storfer24, Townsend}). 
The study of these systems is beneficial because the expected time delays from these systems are useful for constraining the $H_{0}$.  
In this work, we explore the capabilities of the 7-Dimensional Telescope (7DT), a multitelescope system comprising up to twenty 0.5-meter wide-field telescopes \citep{Im2021, g2024, jihoon}, for the initial detection of glSNe (including both Type Ia and CC) under different observing scenarios among this newly identified sample of strong-lens and candidate systems. The 7DT has an inimitable combination of a wide field of view, flexible multi-telescope operations, and transient classification using medium-band filters. These capabilities enable 7DT to discover glSNe.
It is worth noting that due to the limited angular resolution of telescopes, numerous gLSNe remain unresolved \citep{Goldstein}. 
As mentioned in \citet{baglsst}, unresolved glSNe exhibit shorter time delays, which increase the total brightness within a seeing disk. This enhanced brightness offers a distinct advantage for small-telescope arrays with limited sensitivity, thereby enabling the early detection of glSNe.
Also,
the magnification estimates derived from glSNe systems, particularly for glSNe Ia, may offer valuable constraints for modeling the lensing mass distributions in observed systems.
However, the shorter time delay also presents a disadvantage by reducing the precision of time-delay measurements, thus impacting precise estimates of the $H_{0}$.
We select a sample of lensed galaxies and candidates \citep{H2020,H2022,Storfer24} from the footprint of the DESI Legacy Imaging Surveys \citep{dey}. Following the approach presented by \citet{sheu} for generating synthetic glSNe light curves and estimating the expected rate of these systems, we demonstrate the potential of 7DT for the initial detection of glSNe.
Following the initial detection, we propose conducting follow-up observations with powerful telescopes to constrain the $H_{0}$ using both types of glSNe including Type Ia and CC SNe.
We use a model-independent approach \citep{liao1,liao2}, (i.e., without assuming any specific cosmological model) to determine $H_{0}$ 
by anchoring Type Ia SNe from the Pantheon dataset \citep{scolnic} with detected strong lensed systems.

The structure of this paper is organized as follows:
Section \ref{sec2} details the capabilities of the 7DT facility for observation. 
In Section \ref{sec3}, we describe the selection of target fields for glSNe observation, focusing on the spatial distribution of lenses and candidates identified from the DESI imaging survey.
Based on the results of the glSNe simulation, we propose various observing strategies with 7DT.
With respect to the detection rate of glSNe based on the 7DT observing scenario, we provide an estimate of the $H_{0}$ in Section \ref{sec4}. In the concluding section \ref{sec5}, we present a summary of the key findings and discuss the implications of our research.

\section{Observation capabilities of the 7-Dimensional Telescope}
\label{sec2}
The 7DT is a multiple telescope system \citep{Im2021, g2024, jihoon}. This system encompasses twenty 0.5-meter telescopes positioned at the El Sauce Observatory within the Rio Hurtado Valley in Chile. The typical seeing at this site is around 1.5 arcseconds. The angular resolution of the 7DT is primarily constrained by the seeing conditions at the site. Fifteen of the 20 planned telescopes are currently positioned. Each single telescope in this multiple telescope system has a field of view (FOV) 1.27 $deg^2$ (0.92 deg $\times$ 1.38 deg).
  
\begin{figure*}
\centering
\includegraphics[angle=0,width=0.49\textwidth,clip=]{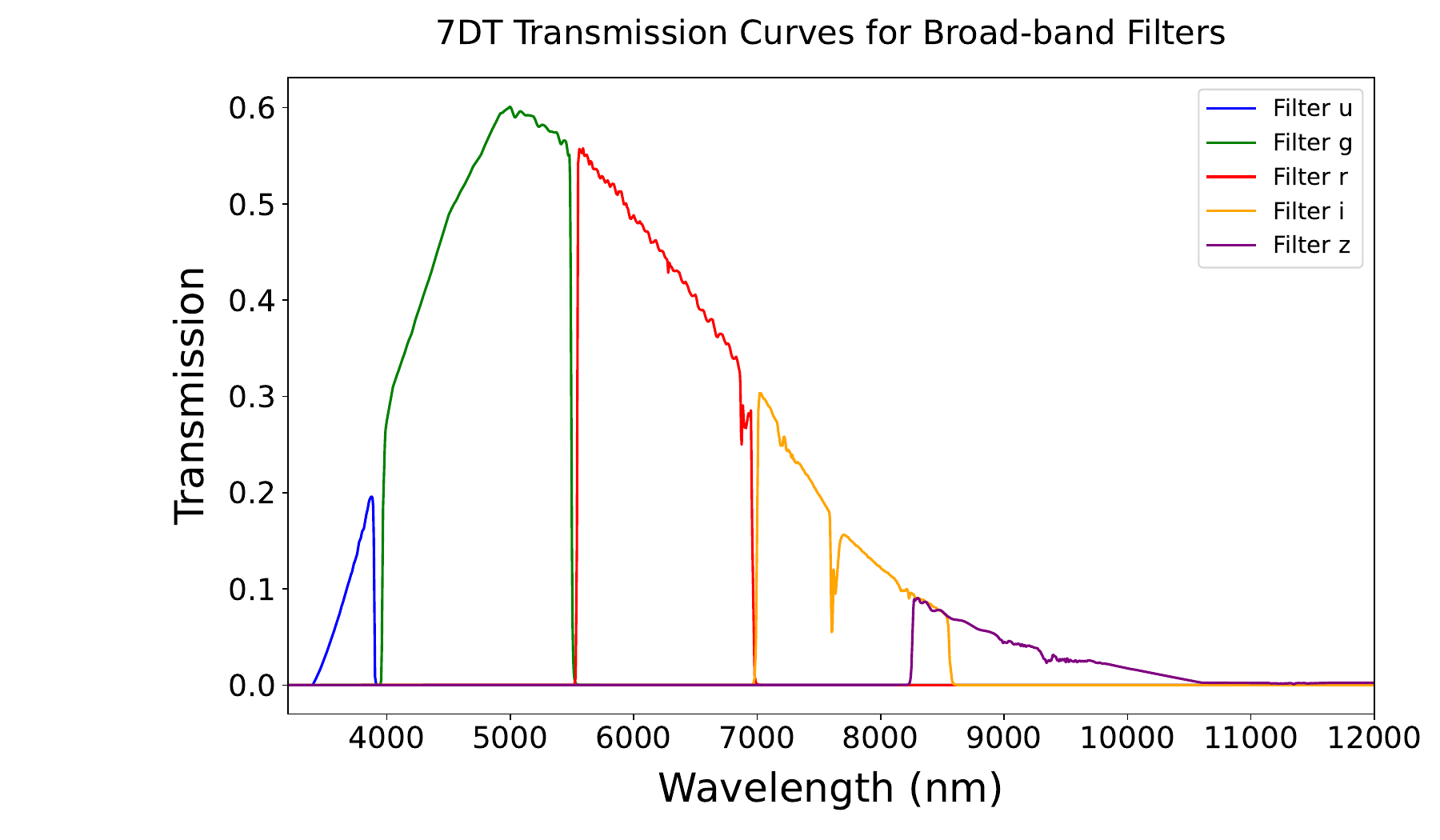}
\includegraphics[angle=0,width=0.49\textwidth,clip=]{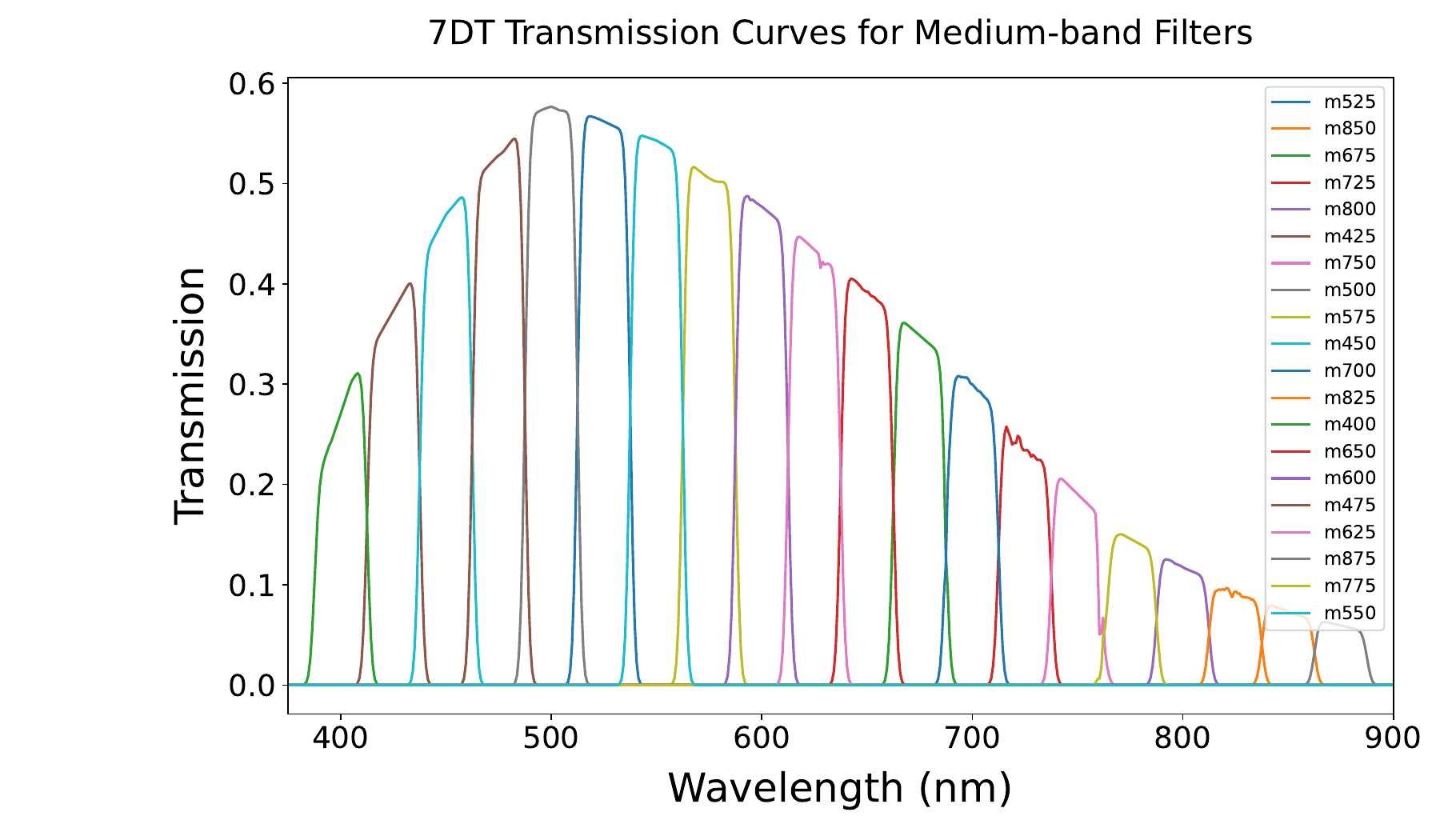}
\caption{The 7DT transmission curves for broad-band and medium-band filters  }
\label{trans}
\end{figure*}

Each telescope in the 7DT system is equipped with Sloan $g$-, $r$-, and $i$-band filters. Additionally, the entire system includes one $u$-band filter and three $z$-band filters. 
The system also includes 40 medium-band filters
characterized by a bandwidth of 25 nm, covering a spectral range from 375 nm  \
to 
900 nm with 12.5 nm gap between them.
At present, 20 medium-band filters are in operation. Their central wavelengths range from 400 nm to 887.5 nm, with 25 nm gaps between them (Figure \ref{trans}).
The 7DT achieves a remarkable combination of features: (i) flexible operation with multiple telescopes, (ii) wide field of view, and (iii) transient classification using medium-band filters. These attributes make it an outstanding discovery instrument for glSNe. 

The 7DT encompasses a variety of operational modes \citep{jihoon}.
These modes include: (1) Spec mode: the 7DT employs various medium-band filters in a single pointing to enable spectral mapping of the sky.
(2) Search mode: the 20 telescopes of the 7DT observe various patches of the sky using broad-band filters. In this mode, the 7DT can survey a vast area of the sky.  The total FOV is approximately (20 $\times$ 1.27) $\sim$ 25 $deg^{2}$.
(3) Deep Observing mode: deep observing mode can operate in several configurations. To streamline operations, we recommend limiting configurations such as 20, 10, 4, or 2 telescopes per pointing. When all 20 telescopes are pointed at the same area using a single filter, the system achieves its highest sensitivity, equivalent to the light-collecting capability of a 2.3-meter diameter telescope.

In the following, we introduce two primary observational strategies implemented with 7DT  \citep{jihoon}:
\begin{itemize}
\item 7-Dimensional Sky Survey (7DS):
7DS is a comprehensive, wide-field time-series survey that employs a multi-object spectroscopy approach with medium-band filters.
The survey observations conducted by the 7DS will include the Reference Image Survey (RIS), the Wide-field Time-domain Survey (WTS), and the Intensive Monitoring Survey (IMS). 

The RIS encompasses an area of 20,000 $deg^{2}$ of the southern sky, excluding the Galactic Plane. Observations are conducted with a uniform cadence and an exposure time of 100 seconds per tile. Each tile is observed three times per visit, totaling a 5-minute exposure time. The total allocated observation time for this survey is 50,000 minutes. The RIS begins in the first year of 7DS observation. This area is observed once to generate reference images, which are utilized for identifying transient events through difference image analysis (DIA). DIA accomplishes this by subtracting the reference image from each individual image (e.g., \citealt{DIA}).

The WTS monitors approximately 1620 $deg^{2}$ of the southern sky with a 14-day cadence over a planned 5-year observational period. Field selection for this survey is currently under discussion, with particular emphasis on regions possessing complementary data, especially those with existing near-infrared photometry.

The IMS covers the AKARI Deep Field South (ADF-S) \citep{akari,akari2}, a $~$ 12 $deg^{2}$ region near the South Ecliptic Pole, with daily observations. The survey allocates a total of 20000 minutes of observation time per year.

\item Target of Opportunity Observation (ToO):
The 7DT utilizes the ToO approach to streamline the detection of various transient phenomena, including the electromagnetic (EM) counterparts of gravitational wave (GW) events.  The same configuration as RIS, specifically using 5-minute exposures with medium-band filters, is applied to this program.
  
\end{itemize}

\section{Exploring observation strategies for the strongly lensed supernovae with 7DT }\label{sec3}
In this section, we introduce observation strategies for glSNe, which have been developed based on the analysis of synthetic glSNe light curves. This development utilizes information from confirmed strong gravitational lenses and candidate lenses identified in the Dark Energy Spectroscopic Instrument (DESI) Legacy Imaging Surveys \citep{H2020,H2022,Storfer24}.

\subsection{Identifying target fields for glSNe observation}\label{rrate}
Recently, \citet{sheu} utilized an archive comprising 5,807 strong lenses and potential candidates identified through the Dark Energy Camera Legacy Survey (DECaLS) to develop a specialized pipeline for searching the glSNe. DECaLS, a key project for the DESI Legacy Imaging Surveys, operates using the Dark Energy Camera mounted on the 4-meter Blanco telescope situated at the Cerro Tololo Inter-American Observatory. This wide-field survey spans a 9000 $deg^{2}$ area of the sky, encompassing both the North Galactic Cap (NGC, declination $<$ 32 deg) and the South Galactic Cap, across the $g$, $r$, and $z-$ bands. 
The lens candidates examined in their search originate from a variety of publications and search efforts. To assess the likelihood that a candidate represents a strong lens system, the criteria used in this paper are similar to those described by \citet{H2022}.
In this paper, we have selected a sample of lens systems and candidates for targeted observation from \citep{H2020, H2022, Storfer24}
\footnote{\url{https://sites.google.com/usfca.edu/neuralens}}, located at declinations below 30 degrees.  
We use the spectroscopic or photometric redshifts of these lens systems and candidates to construct the lens redshift distribution. Then, we adopt the methodology for simulating the redshift of glSNe as outlined by \citet{sheu}. We obtain the source redshift distribution by multiplying the lens redshift distribution by a truncated normal distribution N(2,0.5), with lower bound at 1. 

\begin{figure}
\centering
\includegraphics[angle=0,width=0.49\textwidth,clip=]{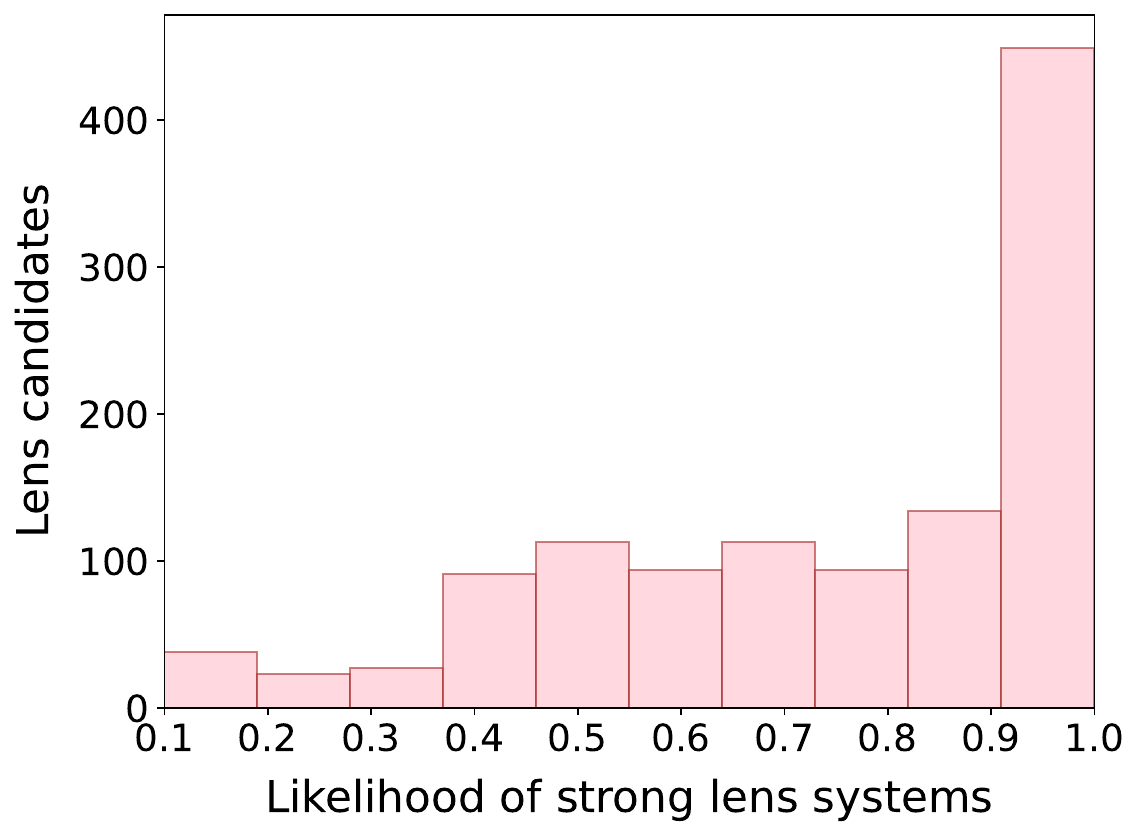}
\caption{Likelihood of lens candidates
}
\label{feas}
\end{figure}

\begin{figure*}
\centering
\includegraphics[angle=0,width=0.32\textwidth,clip=]{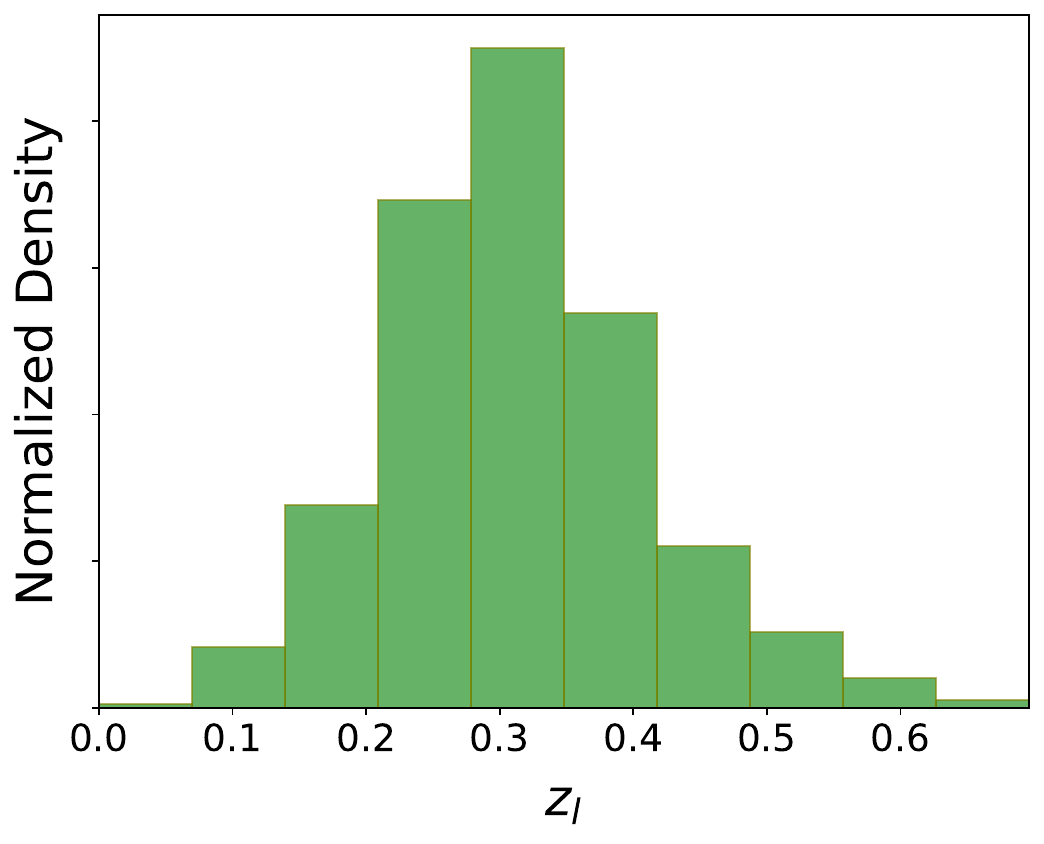}
\includegraphics[angle=0,width=0.32\textwidth,clip=]{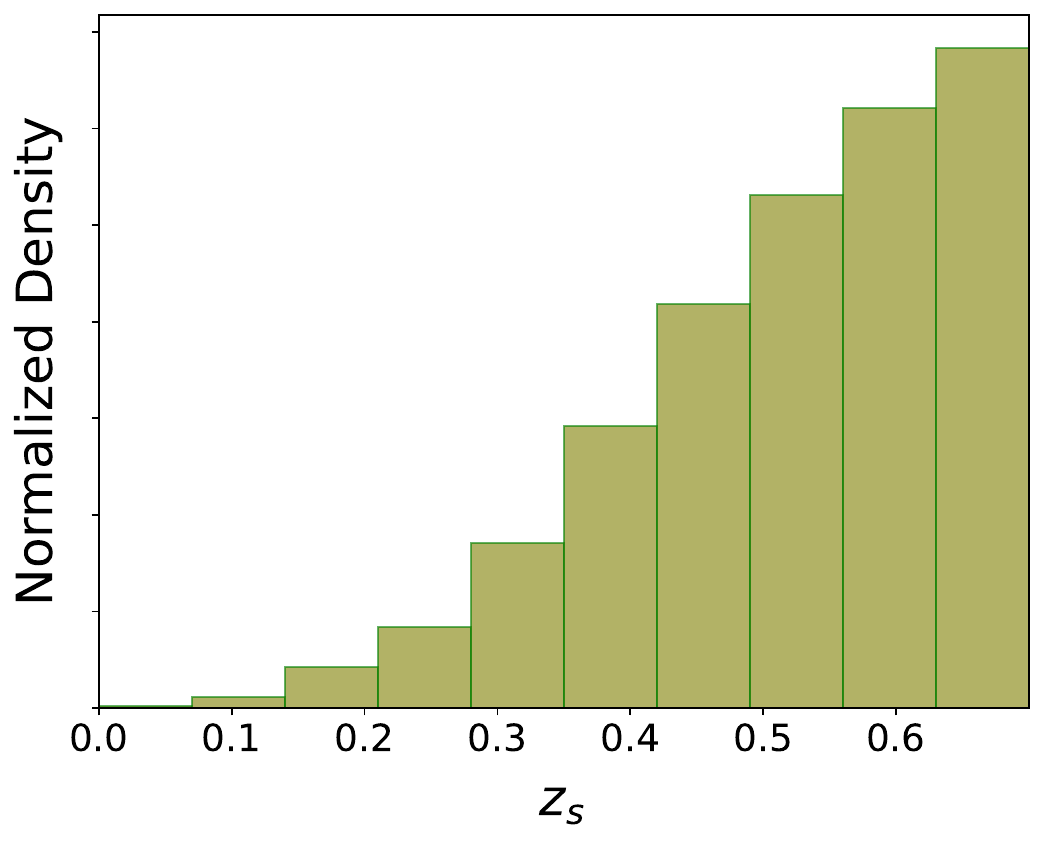}
\includegraphics[angle=0,width=0.32\textwidth,clip=]{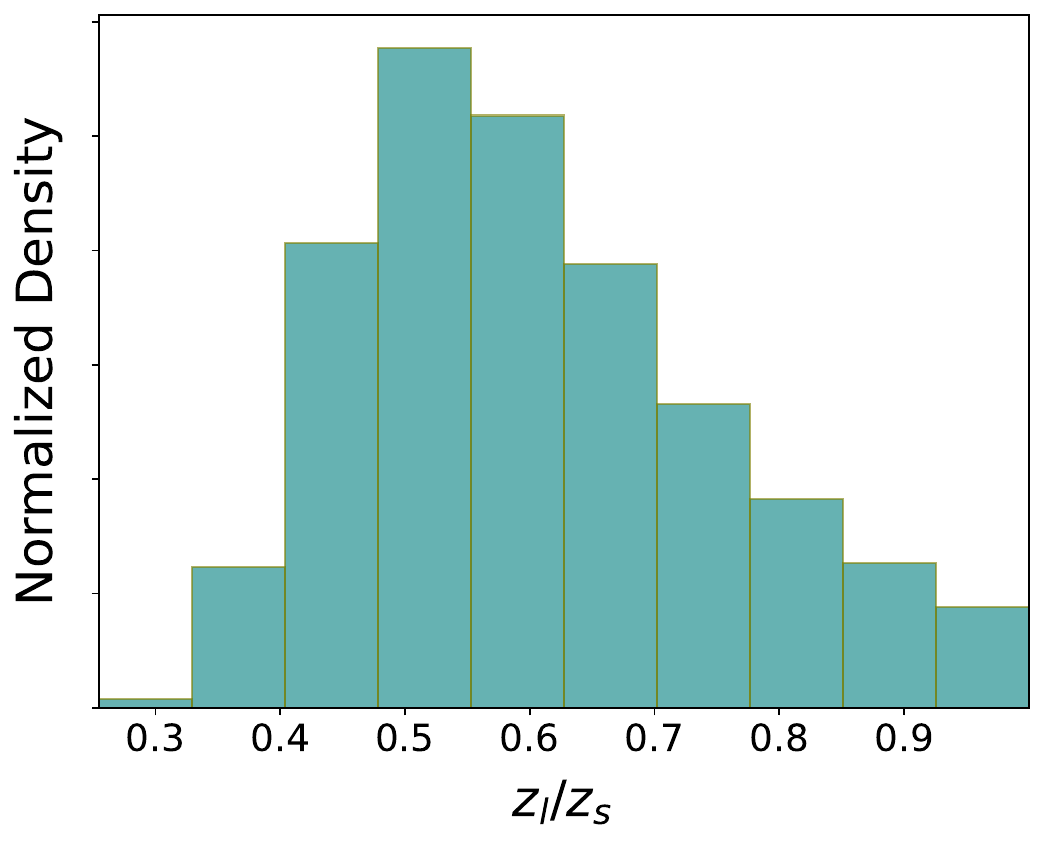}
\caption{Distribution of redshifts for selected lens galaxies/ candidates $z_{l}$, simulated source galaxies $z_{s}$, and the ratio $z_{l}$/ $z_{s}$} 
\label{redsh}
\end{figure*}

\begin{figure}
\centering
\includegraphics[angle=0,width=0.45\textwidth,clip=]{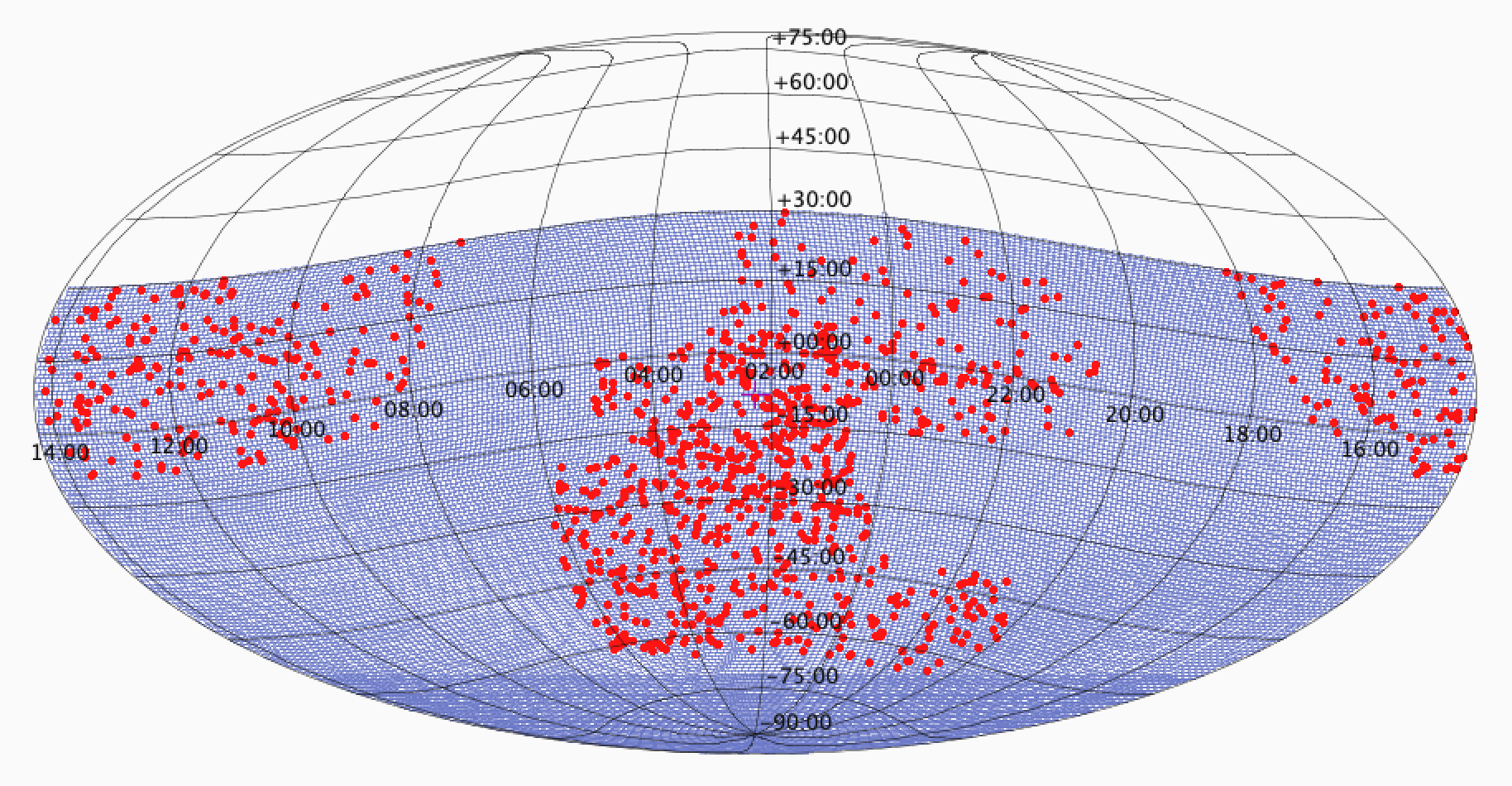}
\caption{7DT skyfootprint (blue tile patterns) and the location of selected lens galaxies/ candidates (red dots). The approximate number of targeted 7DT fields covering the systems is 1125 tiles.
}
\label{foot}
\end{figure}

Furthermore, following the approach described in \citet{sheu}, we estimate the star formation rate (SFR) across different redshifts by fitting a polynomial function to the SFR data provided by \citet{Bell}, \citet{smit}, and \citet{sobral}.
We can calculate the annual CC SNe rate using
\begin{equation} 
R_{CC} = kcc\frac{\text{SFR}}{1+z_s} [\text{yr}^{-1}]
\end{equation}
where the number of CC SNe is $kcc= 0.0068M^{-1}$ 
\citep{shu}
.  
\\
We can estimate the annual rate of SNe Ia through the equation
\begin{equation} 
    R_{Ia} = 0.00084M^{-1}_\odot\frac{ \int^{t(z_s)}_{0.1} \text{SFR}(t(z_s)-t_D)f_D(t_D) \,dt_D }{(1+z_s) \int_{0.1}^{t(z=0)}f_D(t_D)\,dt_D} [\text{yr}^{-1}],
\end{equation}
where $t_{D}$ 
represents the delay time and follows a distribution
$f_D(t_D) \propto t_D^{-1.07} $ \citep{shu}. 
We use the formula from (Equation 2, \citet{sheu}) to calculate the star formation history of lensed sources
($SFR(t(z_s)-t_D)$).
We limit our selection to systems where the glSNe redshift is below 0.7, ensuring they are bright enough to be detected within the sensitivity limits of the telescope. 
We conducted the simulation 100 times to ensure statistically robust and realistic results. 
Figure \ref{feas} illustrates the strong lensing plausibility of these systems. Figure \ref{redsh} presents the redshift distributions of the selected lens systems/ candidates, simulated source galaxies, and the distribution of the ratios of lens redshift to source redshift.\\
We generate sky tiling for 7DT to coordinate glSNe observations. To improve transient identification, we incorporate overlaps between the tiles (Figure \ref{foot}).
Additionally, Figure \ref{foot} displays the spatial locations of the selected lens systems and candidates. Around 1225 lens systems and candidates are distributed across approximately 1125 7DT fields.

In the following sections, we will explore the process of creating synthetic light curves for every system in our simulation. We will also investigate how unresolved glSNe images influence the light curves.
Regarding the output result, we determine the monitoring duration and cadence to give a 
glSNe observing strategy with 7DT.

\subsection{Generating glSNe light curves}
We employ the \texttt{SNCosmo} software \citep{sncosmo} for simulating SNe light curves. We use the \texttt{SALT3} model to generate Type Ia SNe light curves. The critical parameters for this model include redshift, flux normalization ($x_{0}$), color ($c$), and stretch ($x_{1}$) \citep{Kenworthy}. We adopt the same parameter settings as those used in \citet{niki} \footnote{\url{https://github.com/Nikki1510/lensed_supernova_simulator_tool}}.
Specifically, $x_{1}$ is drawn from a skew-normal distribution, $x_{1}\sim\text{$N^{s}$}(a=-8.24,\mu=1.23, \sigma=1.67)$, $c$ follows a  skew-normal distribution, $c\sim\text{$N^{s}$}(a=2.48,\mu=-0.089, \sigma=0.12)$, and the absolute B-band magnitude ($M_{B}$) is sampled from a normal distribution, $M_{B}\sim\text{N}(\mu=-19.43, \sigma=0.12)$.

The \texttt{SNCosmo} package also provides a variety of CC models for SNe, allowing us to choose different models for each SN type. 
We employ various models for our analyses, including  \texttt{Nugent-SN2P} (Type IIP) \citep{iip},  \texttt{SNANA-2004GQ} (Type Ic)
\footnote{\url{http://das.sdss2.org/ge/sample/sdsssn/SNANA-PUBLIC/}}
,  \texttt{V19-2008AQ-CORR} (Type IIb) 
\footnote{\url{https://github.com/maria-vincenzi/PyCoCo_templates}}
,  \texttt{S11-2005HL} (Type Ib) \citep{ib},  \texttt{Nugent-SN2L} (Type IIL) \citep{iip}, and  \texttt{Nugent-SN2N} (Type IIn) \citep{iip}.
These models are characterized by parameters such as redshift ($z$), amplitude, and the time of peak brightness in the B band ($t_{0}$).
We use the distribution of $M_{B}$ and the occurrence rates of various types of CC SNe from Table 1 of \citet{sheu}.
We acknowledge that a more realistic approach would involve randomly sampling from multiple CC models for SNe within each subtype, which represents a limitation of the current study.

For the simulated light curves of glSNe, we consider the effects of dust extinction originating from both the host galaxy and the Milky Way, following the dust extinction model proposed by \citet{fit}. For the host galaxy, we assume a dust extinction characterized by a color excess $E(B-V)_{\mathrm{host}} \leq 0.2$ and a selective-to-total extinction ratio $R_{V}$ in the range $[1.63,\,3.85]$. For the Milky Way extinction, we adopt a standard extinction ratio of $R_{V}=3.1$, with $E(B-V)_{\mathrm{MW}}$ values derived using the \texttt{dustmaps} package \citep{green_dust}, based on the dust reddening map provided by \citet{chiang23}.

\subsection{Unresolved glSNe light curve}
Strong gravitational lensing can produce images with varying magnified fluxes and time delays.
However, because of the angular resolution limitations of ground-based telescopes, which are influenced by atmospheric conditions (seeing), these images may overlap. As a result, they can appear blended or indistinguishable.
When the image separation becomes too small, a unified light curve is recorded \citep{Bag,Denissenya}.
As outlined in \citet{Goldstein}, it is anticipated that more than half of the glSNe detected by LSST will have an angular resolution of less than $1$$\arcsec$. A significant fraction of these systems will exhibit separations below $0.5$$\arcsec$ \citep{baglsst}. Given that the typical seeing for LSST is $0.7$$\arcsec$ in the $r-$ band \citep{lsstp}, a considerable portion of the strong lensing systems will likely be unresolved by this wide-field survey.
\citet{baglsst} reported that unresolved systems have shorter time delays compared to resolved systems. Their findings indicate that, for unresolved systems in the LSST, the median time delay is 2.03 days. And only around 10\% of these systems have time delays over 10 days. The findings presented in Figure 11 of \citet{ztf} for the Zwicky Transient Facility (ZTF), with a spatial resolution of 2$\arcsec$ for this telescope, reveal that most events have angular separations below 1$\arcsec$. Furthermore, the majority of these events exhibit time delays of less than 10 days, with a median delay of approximately 5 days.

Following the methodology outlined in \citet{sheu}, we assume that each lensing system produces either 2 or 4 lensed images with probabilities of 0.7 and 0.3, respectively, consistent with the results from \citet{Oguri2010}. For each system, \citet{sheu} sampled the magnifications from a log-normal distribution with a mean of 1.5 and a standard deviation of 0.35, resulting in an expected magnification of 4.765. Also, they sample the time delays between the lensed images for each system from a normal distribution, $N(36, 4)$ days, as described in \citet{Craig}. 
Inspired by the findings of \citep{baglsst, ztf}, we refine the time-delay and magnification distributions in our simulation. 
We select the magnification as a log-normal distribution with mean= 1.07 and standard deviation= 0.465.  
We sample the time-delay distribution from an exponential distribution
\footnote{We select this distribution based on Fig. 3 of \citet{baglsst} and the time-delay histogram in Fig. 11 of \citet{ztf}}
with a mean of 6.83. 
The mean and standard deviation are selected based on the values provided in Table 3 of \citet{ztf}
\footnote{We convert the magnification value to a logarithmic scale}.
We conduct our simulation in two steps. First, we assume that approximately $50\%$ of the images in a system are completely unresolved. In the second step, we run the simulation under the assumption that $90\%$ of the images in a system are completely unresolved.
\begin{table}[ht]
\centering
\caption{5$\sigma$ depth magnitudes across various filters for the 7-DT in 300s exposure times. }
\label{filter_m}
\begin{tabular}{c|c}
\hline
\hline
\textbf{Filter (medium-band)} & \textbf{5$\sigma$ Depth (AB mag)} \\
\hline
$m_{425}$ & 20.88 \\
$m_{450}$ & 20.91 \\
$m_{475}$ & 20.94 \\
$m_{500}$ & 20.95 \\
$m_{525}$ & 20.90 \\
$m_{550}$ & 20.70 \\
$m_{575}$ & 20.72 \\
$m_{600}$ & 20.60 \\
$m_{625}$ & 20.40 \\
$m_{650}$ & 20.42 \\
$m_{675}$ & 20.32 \\
$m_{700}$ & 20.18 \\
$m_{725}$ & 19.85 \\
$m_{750}$ & 19.74 \\
$m_{775}$ & 19.35 \\
$m_{800}$ & 19.19 \\
$m_{825}$ & 18.97 \\
$m_{850}$ & 18.76 \\
$m_{875}$ & 18.36 \\
\hline
\end{tabular}
\end{table}

It is important to note that, for simplicity, we ignore the microlensig effect due to
stars in the lens galaxy or in the foreground
lens galaxy. Microlensing can affect the observed brightness of glSNe, either magnifying or suppressing their brightness. Although brightness suppression occurs more frequently than magnification \citep{gold2018, niki}.
Microlensing also impacts unresolved sources differently from resolved ones. For unresolved sources, the observed flux is a sum of the individual fluxes from multiple images, each of which may be intrinsically fainter compared to those of resolved sources. Although microlensing affects each individual image differently, the summed flux of unresolved images is often still above the detection threshold, effectively reducing the overall impact of microlensing on peak brightness measurements. As a result, incorporating microlensing into analyses would likely increase the predicted ratio of unresolved to resolved SNe detections \citep{baglsst}. Furthermore, microlensing introduces substantial uncertainties in time-delay measurements \citep{gold2018}, particularly for glSNe with inherently short delays (e.g.,  \citealt{goobar, Huber2019}). While acknowledging the importance of microlensing, we explicitly state that modeling its effects is beyond the scope of this paper and constitutes a caveat of our analysis.

\subsection{Observing scenarios}
In this subsection, we present various observing strategies with 7DT and calculate the annual detection rates for each strategy. 
\subsubsection{7-Dimensional Sky Survey (7DS):}
\begin{figure*}
\centering
\includegraphics[angle=0,width=0.7\textwidth,clip=]{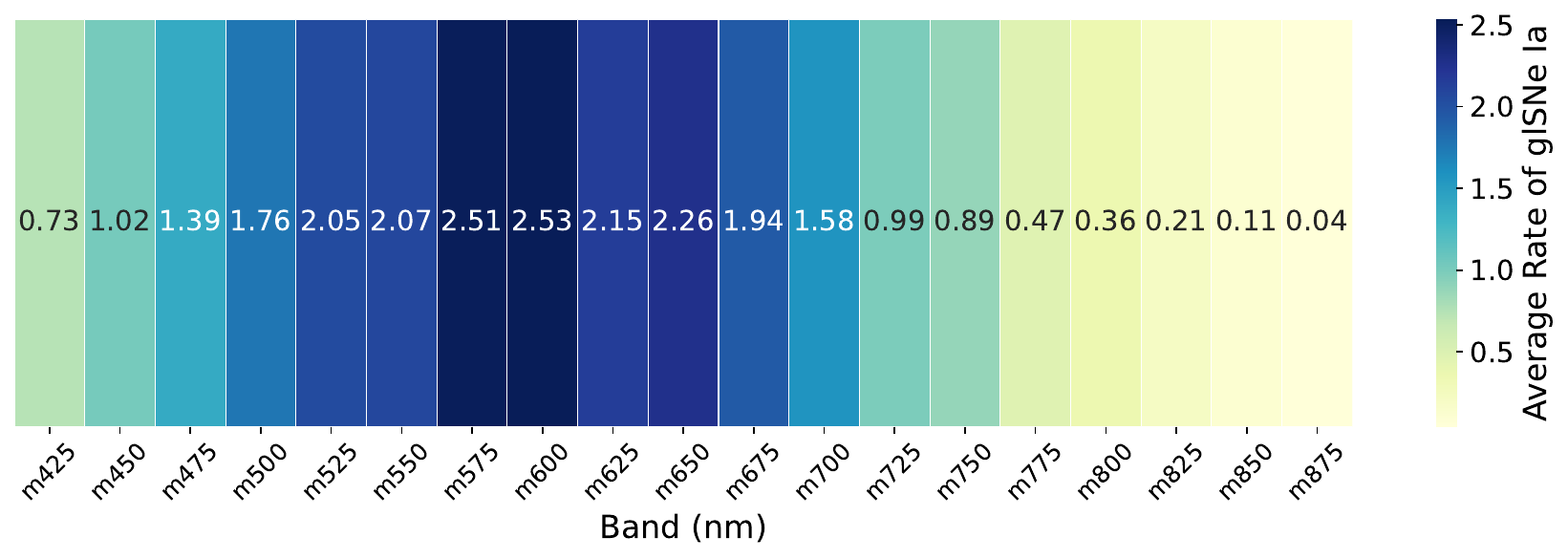}  
\caption{Average detection rate of glSNe Ia for different 7DT medium-band filters and source redshifts z$_{s}$ $\le$ 0.6, assuming $90\%$ of systems are unresolved.}
\label{medium_rate}
\end{figure*}

 The 7DS conducts the WTS as a core program in spec mode for a 5-year observation period. The advantage of observing in different medium-band filters is transient classification.
The exposure-time for this program set to 300s, The 5$\sigma$ depth of the 7DT for the medium-band filters at 300s is listed in Table \ref{filter_m}.  Note that the depths given for various exposure times are simulated values based on ideal conditions, including 1.5$\arcsec$ seeing, gray moon phase, and marginal target altitude.
To optimally determine the cadence for glSNe observations, we calculate the control time of mock light curves in our simulations. The control time is determined as follows:
\begin{itemize}
\item For systems with resolved images: We consider the first bright image in the system. The control time is defined as the width of the light curve of this image at a magnitude equal to the observing depth of the telescope.
\item For systems with unresolved images: The control time is determined as the width of the combined light curve at a magnitude equal to the observing depth of the telescope.
\end{itemize}
Since WTS conducts a relatively shallow survey, only the brightest objects can reach our detection limits. Considering the intrinsic brightness of different types of SNe, our simulations indicate that SNe Ia (that are intrinsically much brighter than other SNe types) are the most likely to be observed. While some highly magnified or intrinsically bright CC SNe could still be detectable, their expected occurrence is lower.\\
The control time distribution of glSNe Ia, assuming that $90\%$ of the systems have completely unresolved images, across various medium-band filters is presented in Figure \ref{medium_T}. 
These findings highlight that, as discussed in Section \ref{sec2}, the 14-day cadence proposed for this program is well-suited for glSNe Ia observations.
We estimate the annual detection rate of glSN Ia events based on the depth of observations conducted with the 7DS-WTS program in spec mode. For this estimation, we use the formation rate formulation of glSNe Ia presented in \ref{rrate} as a first-order approximation. A system is considered detectable with 7DT if the peak brightness of at least one image within the system surpasses the observation depth. Figure \ref{medium_rate} illustrates the detection rates of glSNe Ia, assuming that $90\%$ of the systems have completely unresolved images. We expect to observe up to 2.53 events per year for systems with a redshift $<$ 0.6 as part of the 7DS-WTS program. 
However, based on simulation results, systems with redshift in the range $0.6 < z_{s} < 0.7$ are not detectable by the WTS program.
It is important to highlight that the expected detection rates depend on observing conditions, such as moon phase and seeing, and can vary accordingly.
\subsubsection{7DT target program for glSNe observations}
We recommend implementing the glSNe observations as a program designed to target specific regions of the sky. This program can operate in search mode or deep mode as a targeted survey utilizing a broad-band filter. The glSNe exhibit redder colors, indicating that observations in the near-infrared are more suitable. However, due to the lower sensitivity of the 7DT in the $i$- and $z$-bands, we focus on detection of glSNe in the $r$-band.
For this program, we recommend a monitoring duration from 7 to 14 days. 
This monitoring cadence is derived from the analysis of the control time distributions generated from mock light curves in the $r$-band filter, at depths of 21.02 magnitudes (with an exposure time of 60 seconds) and 22.04 magnitudes (with an exposure time of 360 seconds). Figure \ref{f1} (upper panel for glSNe Ia) and Figures \ref{b1} and \ref{fcc} (for glSNe CC) illustrate the control time distributions corresponding to $50\%$ and $90\%$ of systems with completely unresolved images.
\\
This targeted survey is categorized into two groups based on the depth of observation and required exposure times to achieve certain magnitudes: (1) the shallow observation, which enables us to reach a magnitude of $\le$ 21.02 and (2) the deep observation, designed to seek a deeper insight with a magnitude of $\le$ 22.04 mag.
\begin{itemize}
\item shallow observation: Each telescope in the 7-DT array can achieve a magnitude of 21.02 with a one-minute exposure. With an observing duration of 7 hours per night and accounting for approximately 10$\%$ overhead (including 10-second slewing, 100-second auto focus, 5-second image readout, and 5–10 seconds for filter changes and focusing), we can cover all target tiles ($\sim$ 1125) within 2 hours.
By observing in a single filter, we can effectively identify transients by employing reference images obtained from the Reference Image Survey (RIS).
\item Deep observation: 
As previously discussed, various configurations can be applied to deep-mode observation. Given that a single 7DT telescope can achieve a 22.04 magnitude with a 6-minute exposure time in the $r-$ band, we recommend utilizing two telescopes per pointing, with each tile observed for 3 minutes. However, it is important to note that increasing the number of telescopes per tile will lead to higher overhead times due to the need for telescope adjustments. In this case, for 7-hour observations per night, all target tiles can cover to depth 22.04 mag. We conduct observations using at least one single filter ($r-$ band). 
\\
It should be noted that the expected exposure time can vary depending on conditions such as the moon phase and seeing. The suggested exposure times, that is, 1 minute for a depth of 21.02 and 6 minutes for a depth of 22.04, are based on simulations under ideal conditions. When observing under less favorable conditions, longer exposure times are required. Consequently, to cover all target tiles in real observations, additional observing time may be necessary. 
\end{itemize}
If a transient is detected, follow-up observations using different medium-band filters should be performed in the same field to verify the characteristics of the transient.
\\
We provide a rough estimation of the glSNe detection rate in the $r$-band. 
For this estimation, we adopt the formation rate formulation of glSNe Ia and CC outlined in \ref{rrate} as a first-order approximation. 
Figures \ref{f1} (lower panel for glSNe Ia) and Figures \ref{c1} and \ref{fcc-rate} (for glSNe CC) display the detection rates at depths of 21.02 and 22.04 magnitudes, representing $50\%$ and $90\%$ of systems with completely unresolved images.  
\\
It is important to emphasize that the LSST provides significantly more advanced observational capabilities compared to the single r-band 7DT target program for glSNe observations. However, since not everyone has access to the LSST and many astronomers rely on smaller telescopes, our motivation for this observing strategy is to demonstrate that even modest facilities can still make meaningful contributions to cosmology.
\subsection{Integrating 7DT with LSST for glSNe observation}
The LSST
\footnote{\url{https://www.lsst.org/}}
is an 8.4-meter ground-based telescope located on Cerro Pachón in north-central Chile. The LSST is designed to conduct a multi-band imaging survey, covering approximately 20000 square degrees of the sky with a 9.6 square-degree field of view \citep{lsstp}.
Several studies \citep{Goldstein, niki, Rydberg, Wojtak} have estimated that LSST could detect on the order of tens to hundreds of glSNe annually. The expected number of these detections depends on the observing strategy, which varies with factors such as cadence, depth, filter configuration, and sky coverage. 
The LSST observing scenario is expected to implement rolling cadence strategies approximately 1.5 years after the survey begins. In this approach, the LSST Wide Fast Deep (WFD) footprint is divided into multiple regions that cycle between high and low observational cadence across survey years. The rolling cadence results in a non-uniform distribution of visits across seasons. In some seasons, certain regions of the WFD footprint receive more than the typical number of visits (high cadence), while in other seasons, those same regions receive fewer than average (low cadence).  
During low-cadence seasons, each field typically receives around 25 visits. The season length remains approximately the same in both low- and high-cadence seasons, typically spanning about 180 days
\footnote{\url{https://survey-strategy.lsst.io/baseline/wfd.html\#wfd-rolling-cadence}}.\\
\citet{Huber2019} and \citet{niki} investigate the impact of rolling cadence strategies on the number of glSNe detections with LSST. 
 
As shown in Figure \ref{foot}, the distribution of glSNe systems and candidates spans both high- and low-cadence regions within the WFD footprint of LSST. 
We expect the 7DT strategy to have a higher impact if it targets glSNe fields located in low-cadence regions during the rolling-cadence phase of LSST, although we do not simulate this observing strategy in this paper.
The 7DT observes these target fields with broad-band filters. If a transient candidate is detected, targeted follow-up observations with a set of medium-band filters are performed in the same field to verify and characterize the transient in detail.  
This allows 7DT to provide higher-cadence monitoring in those under-sampled areas, enabling earlier detections and improved transient classification through its dedicated medium-band filter set. 
\section{Cosmology: $H_{0}$ constraint }\label{sec4}  
In this section, we apply a model-independent method, as presented by \citet{liao1,liao2,Li}, to constrain the $H_{0}$. This is achieved through anchoring the relative distances of SNe Ia from the Pantheon dataset \citep{scolnic} with time-delay distance measurements of glSNe. 
In this approach, by combining a large statistical sample of Type Ia SNe, which are insensitive to $H_{0}$, with a smaller sample of strongly lensed systems that are sensitive to $H_{0}$, we can estimate $H_{0}$ without relying on any specific cosmological model \citep{liao1}.
These glSNe are anticipated to be initially detected with the 7DT and subsequently followed up with high-powered telescopes.

The idea is to forecast the ability to constrain $H_0$ by generating a mock time-delay distance dataset alongside a mock SNe Ia standard candle dataset. We then use Gaussian process (GP) regression to generate realizations of $H_0D^L$ from the SNe Ia standard candle dataset, which are then anchored by the time-delay distance dataset.  This works by evaluating the GP reconstructions of $H_0D^L$ at the mock strong lens redshifts, for each strong lens system, then turning $H_0D^L$ into $H_0D_{\Delta t}$.  With a value of $H_0$, we can then compare  the ``model'' time-delay distances to the mock ``data'' time-delay distances (evaluate a likelihood), for each realization of the GP reconstruction. The posterior on $H_0$ is then just marginalizing over the GP realizations. These sort of model-independent constraints are important since, in these high-precision regimes, the assumption of a background cosmological model can bias the inference of parameters~\citep{GW_bias,GW_GP}.

We utilize GP regression 
\footnote{\url{https://zenodo.org/records/999564}}
on the Pantheon SNe Ia dataset \citep{PP,scolnic}.
We use this method to generate 1000 reconstructions of the unanchored luminosity distance ($H_{0}$- independent quantity denoted as $H_{0} D^{L}$) from the SNe Ia data (Figure \ref{GP}). 
Then, we calculate the unanchored angular diameter distance, represented by $(H_{0} D^{A})$, using the formula $H_{0} D^{A} = \frac{(H_{0} D^{L})}{(1+z)^2}$ \citep{Hogg}. 
For each system identified in Section \ref{sec3}, we compute $1000$ $H_{0} D^{A}$ at the the lens and source redshifts, denoted as 
$H_0D_{\rm{l}}$ and $H_0D_{\rm{s}}$ respectively.
Subsequently, we determine the time-delay distances ($D_{\Delta t}$) for each system in our gravitational lensing simulation. For every identified system, we calculate $1000$ values of $H_{0} D_{\Delta t}$ using the equation \citep{Refsdal264,Schneiderl,suyul}: 
\begin{equation} \label{eqq1}
H_0D_{\Delta t,j,i}=\,(1+z_{\rm{l},i})\frac{(H_0D_{\rm{l},j}(z_{l,i}))(H_0D_{\rm{s},j}(z_{\rm{s},i}))}{(H_0D_{\rm{ls},j}(z_{\rm{l},i},z_{\rm{s},i}))},
\end{equation}
where l and s represent lens and source respectively and $D_{ls}$ represents the distance between the lens and the source \citep{Weinberg}. Also, $j$ is the index of the realization of the GP reconstruction and $i$ is the index of each of the $N$ mock strong lens for the different cases.

In the next step, we calculate the likelihood  for glSNe systems as follows \cite{Li} for each realization of the GP reconstruction:
\begin{equation}
\ln{\mathcal{L}}_{D_{\Delta t},j}(H_0,j)\,=\,-\frac{1}{2}\sum_{i=1}^{N} \left(\frac{ \frac{1}{H_0}H_0D_{\Delta t, j,i}-D_{\Delta t,i}^{\rm{sim}}}{\sigma_{D_{\Delta t,i}}}\right)^2.
\end{equation}
$D_{\Delta t,i}^{\rm{sim}}$ is the mock time-delay distance data. This mock data is calculated by taking a flat $\Lambda$CDM model with $H_0=70$ km/s/Mpc and $\Omega_m=0.3$ and evaluating time delay distances at redshifts sampled from our procedure defined in previous sections. Then 10\% noise is added to these mock time-delay distances, which we estimate to be a reasonable uncertainty associated with the time-delay distances ($\sigma_{D_{\Delta t,i}} = 0.1 D_{\Delta t,i}^{\rm{sim}}$).
We vary the value of $H_{0}$ in the range $[60, 80]$ km/s/Mpc with 121 steps.
Finally, we calculate the posterior for $H_0$ by marginalizing over the realizations of $H_0D^L$ GP reconstructions:
\begin{equation}
    P(H_0) = \sum_{j=1}^{1000} \exp( \ln{\mathcal{L}}_{D_{\Delta t, j}}(H_0,j)).
\end{equation}

Table \ref{hubble} presents the best-estimated values of $H_{0}$ for the expected number of glSN systems, including:  
(1) 7 glSNe Ia, (2) 7 glSNe CC, and (3) the combined sample of glSNe Ia and CC detected through the 7DT deep-target program. It also includes the $H_{0}$ estimation for two glSNe Ia detected through the 7DS-WTS program. 
\section{Conclusions}\label{sec5}
Due to the angular resolution limitation of telescopes, a significant portion of glSNe systems remain unresolved. As demonstrated in \citep{baglsst}, unresolved systems have shorter time delays relative to  resolved systems. The light curves of unresolved systems are the result of the combined contributions from the individual images of the system. These shorter time delays lead to an increased brightness in the summed light curve, which is advantageous for
array of small telescopes 
with low limiting magnitudes, enabling the initial detection of glSNe. 
However, shorter time delays pose a disadvantage for precise time-delay measurements, thereby impacting the estimation of $H_{0}$ precisely.
In this paper,
we examine the capability of the 7DT to discover glSNe under different observing strategies. 
We utilized a catalog of strong lens systems and candidates observed by the DESI Legacy Imaging Surveys \citep{dey}. By using the lens redshift distribution provided in this catalog, along with the formula presented by \citet{shu, sheu}, we conducted simulations to generate various characteristics of lensed systems, including
source redshifts, the number of lensed images, magnifications for lensed systems, time delay between images, and the rates of Type Ia and CC SNe. Subsequently, we create synthetic light curves for each of the simulated SNe, taking into account an effect of dust from the host galaxy and the Milky Way on these curves. As mentioned by \citet{Goldstein}, the majority of lensed systems remain unresolved due to the resolution limitations of telescopes. This factor impacts the light curves. To address this, we execute our code twice: initially under the assumption that the generated images of $50\%$ of the systems are unresolved, and subsequently assuming that $90\%$ are unresolved. 
\\
\indent Based on the simulation outcomes, we anticipate detecting up to 2 glSNe Ia with the 7DS-WTS program and up to 7 glSNe Ia and 7 CC events under the 7DT target program. Then we assume that the glSNe initially detected by the 7DT will be followed up with more powerful telescopes.
Furthermore, we propose a collaborative observing strategy that combines the capabilities of the 7DT and LSST for glSNe observation.
In the next step,  
we perform a model-independent analysis, free from any assumption about cosmological models, to constrain the $H_{0}$ using GP regression by anchoring the SNe Ia from Pantheon dataset with time-delay distances from detected glSNe both Type Ia and CC.
Our model-independent results yield  $H_{0}$= 71.4 $\pm$ 5.1 km/s/Mpc for 2  glSNe Ia detected with the 7DS-WTS program, and  $H_{0}$= 70.03 $\pm$ 1.9 km/s/Mpc for the 7 glSNe Ia and 7 glSNe CC detected under the deep-targeting scenario proposed for the 7DT observing program.

\begin{acknowledgements} 
We thank Hyeonho Choi for providing information regarding the 7DT overhead time.
E.K, A.S and H.M.L are supported by the National Research Foundation of Korea 2021M3F7A1082056.
GSHP and MI acknowledge the support from the National Research Foundation of Korea (NRF) grant, No. 2021M3F7A1084525, funded by the Korea government (MSIT).

\begin{figure*}
\centering
\textbf{7DT: SNe Ia}\par\medskip  
\begin{minipage}{\textwidth}
 
  \includegraphics[angle=0,width=0.5\textwidth,clip=]{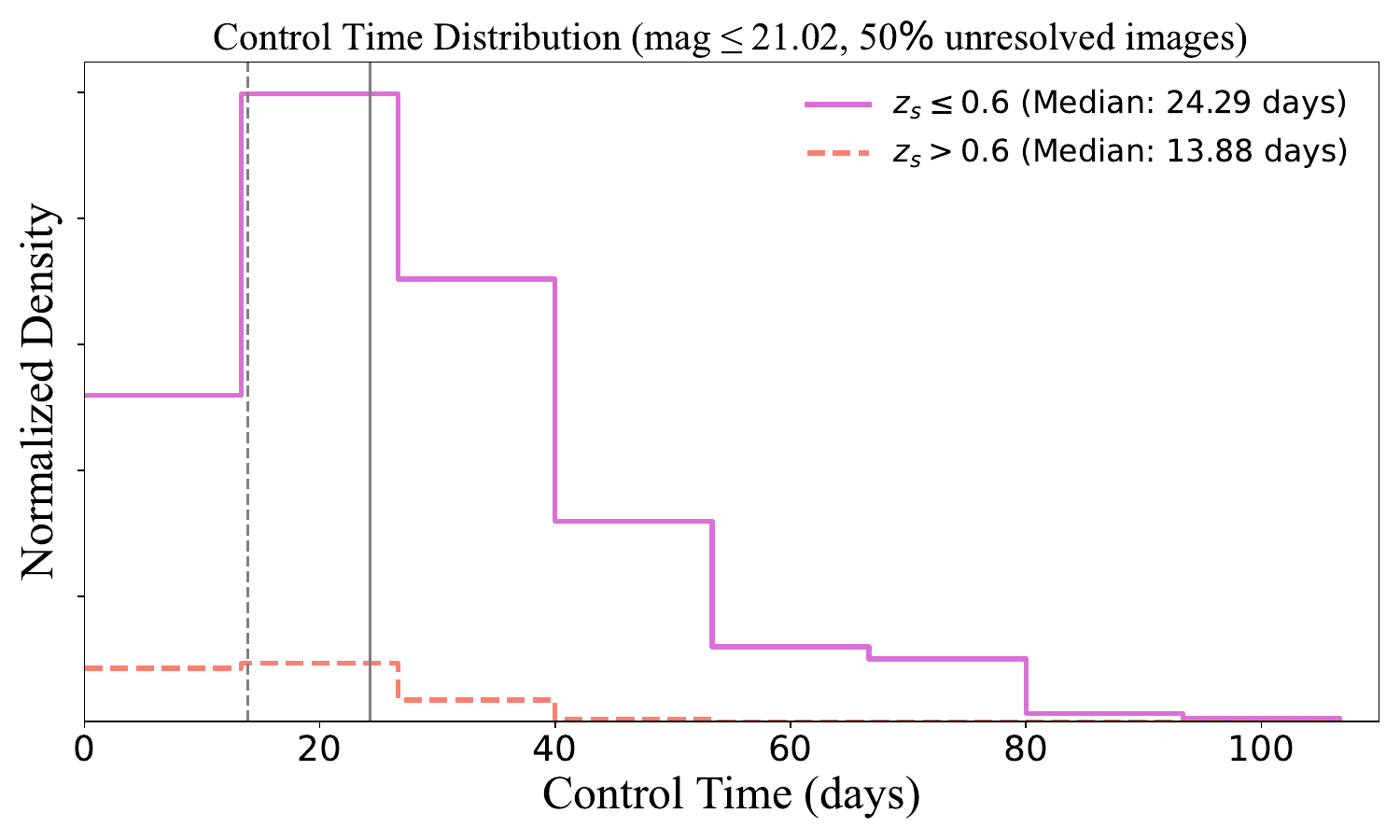}
  \includegraphics[angle=0,width=0.5\textwidth,clip=]{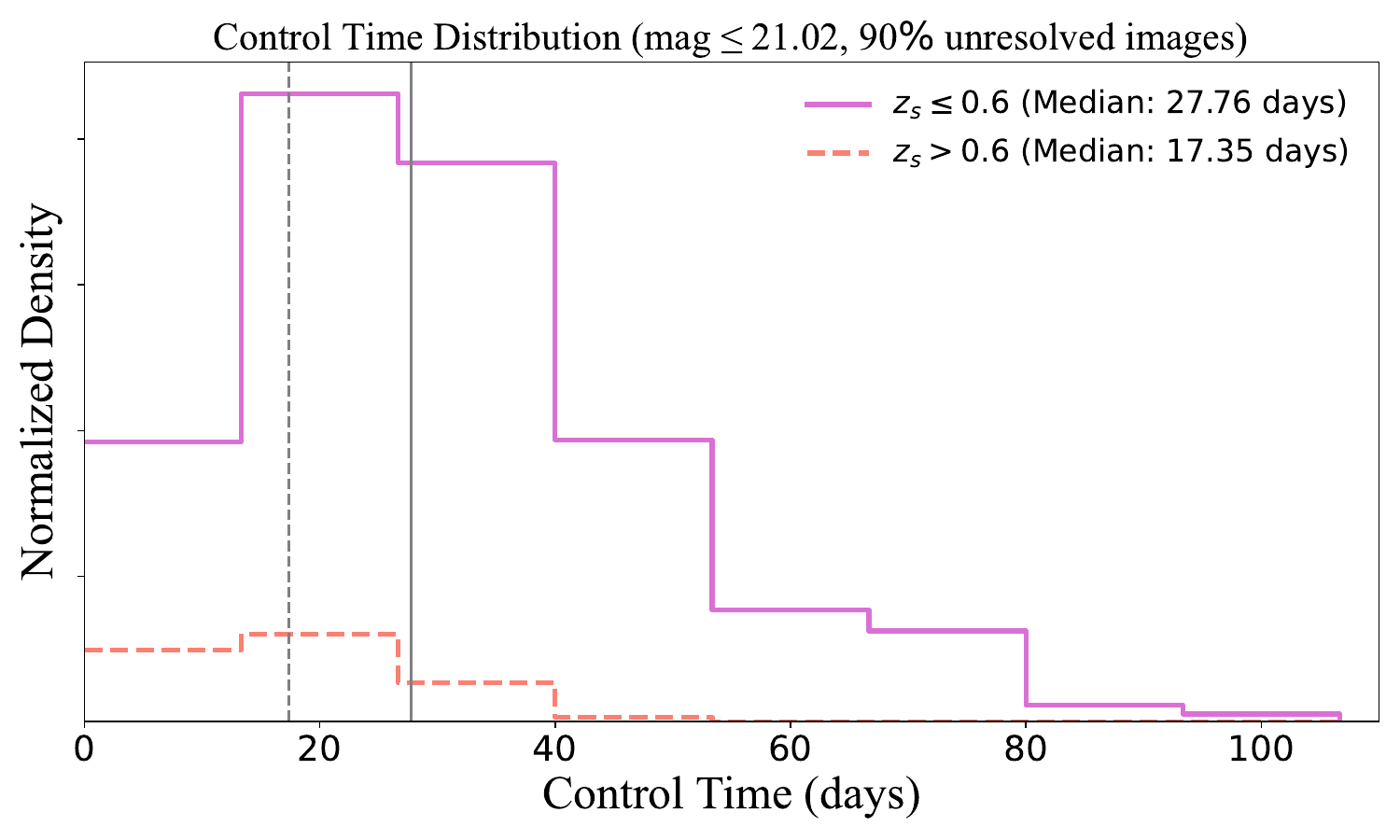}
\end{minipage}
\begin{minipage}{\textwidth}
 \includegraphics[angle=0,width=0.5\textwidth,clip=]{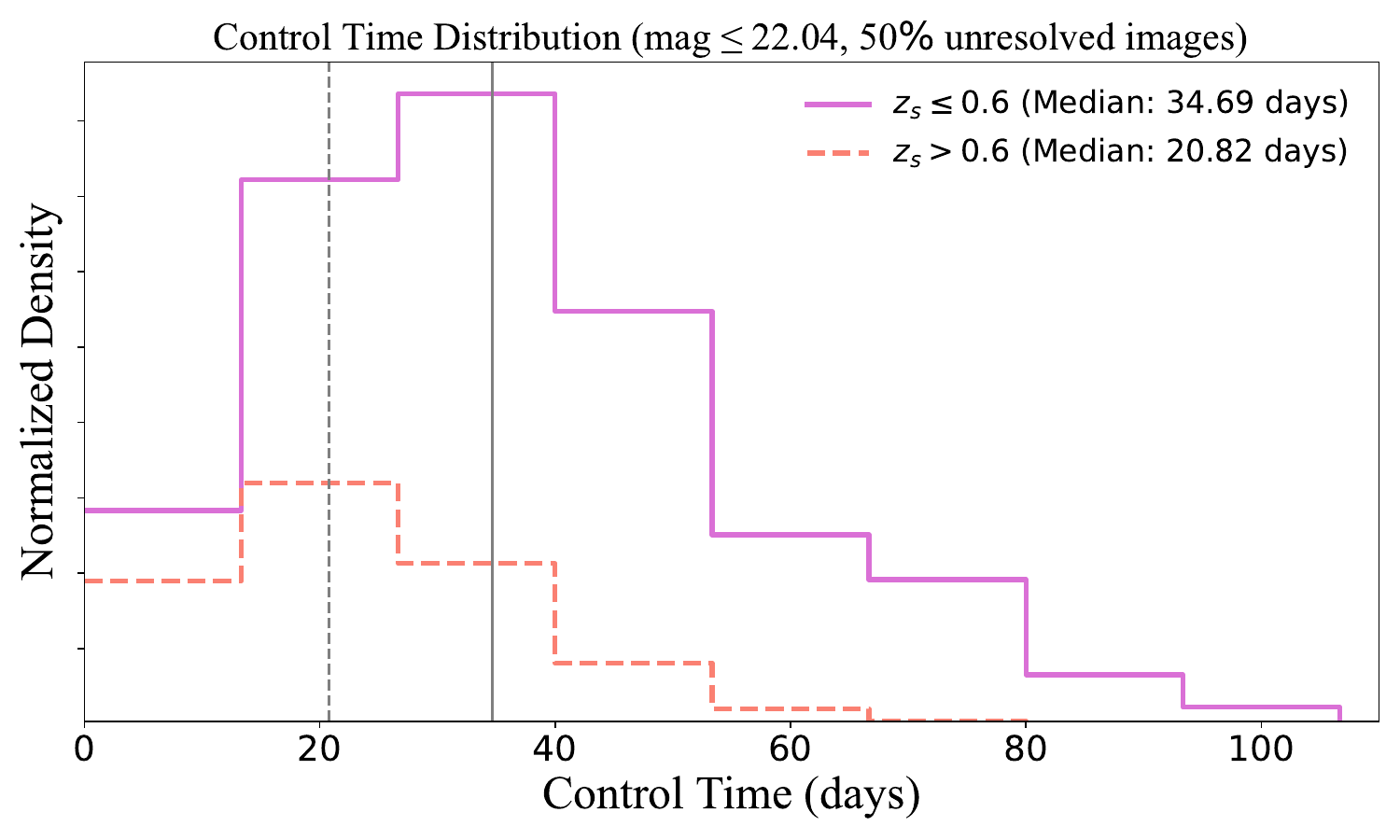}
  \includegraphics[angle=0,width=0.5\textwidth,clip=]{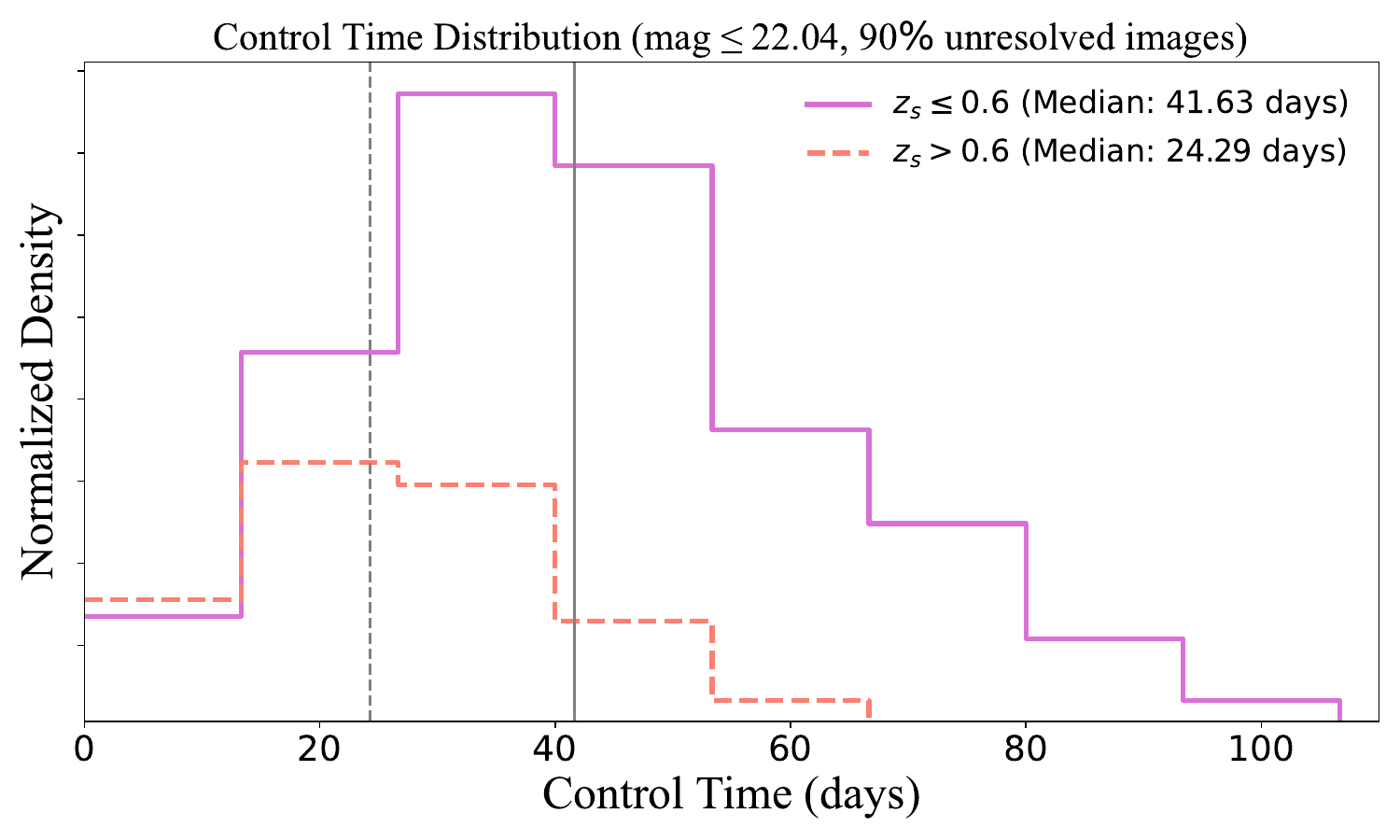}
\end{minipage}
\vspace{0.2cm}
\begin{minipage}{\textwidth}
\centering
\includegraphics[angle=0,width=0.406\textwidth,clip=]{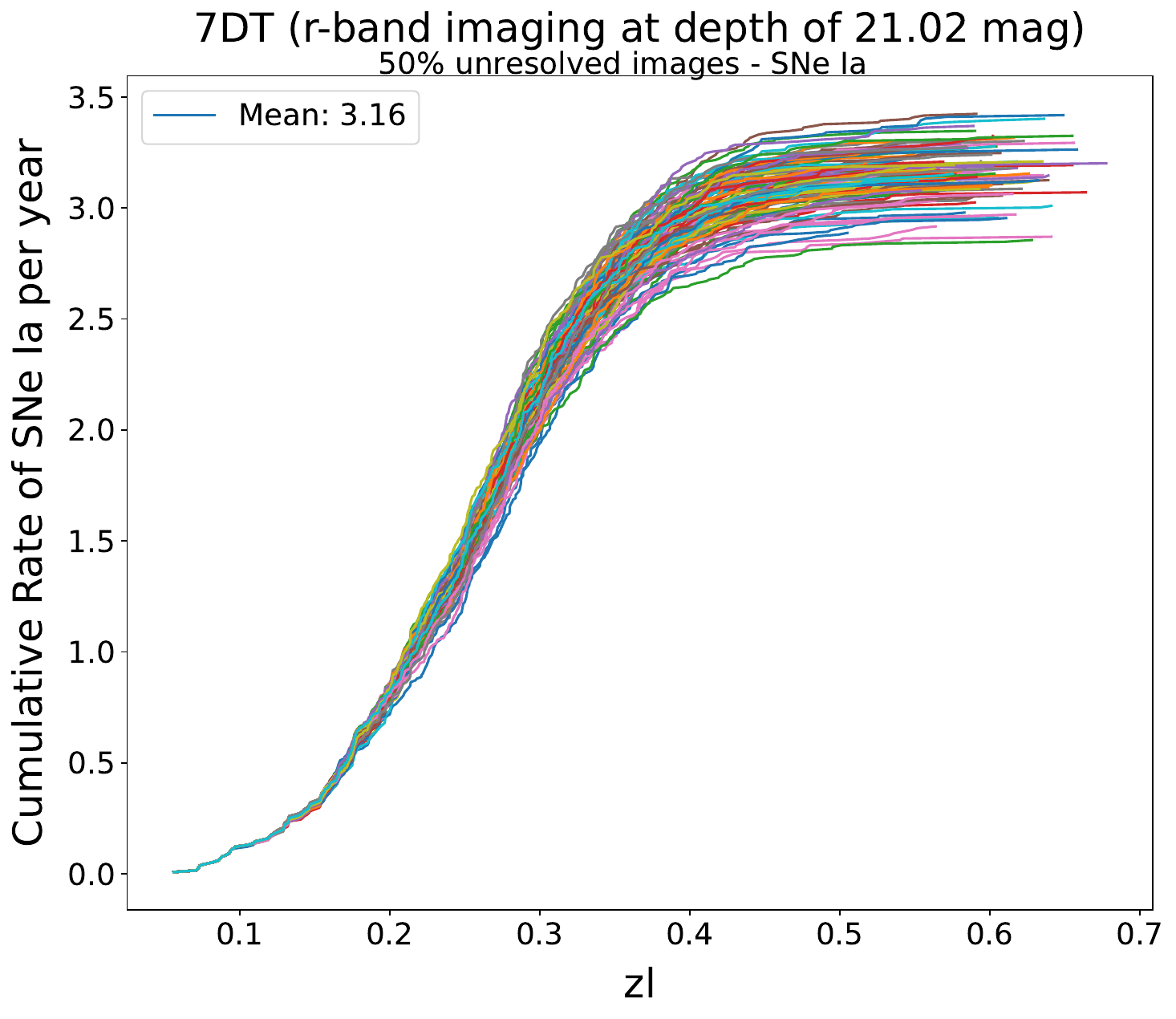}
\includegraphics[angle=0,width=0.4\textwidth,clip=]{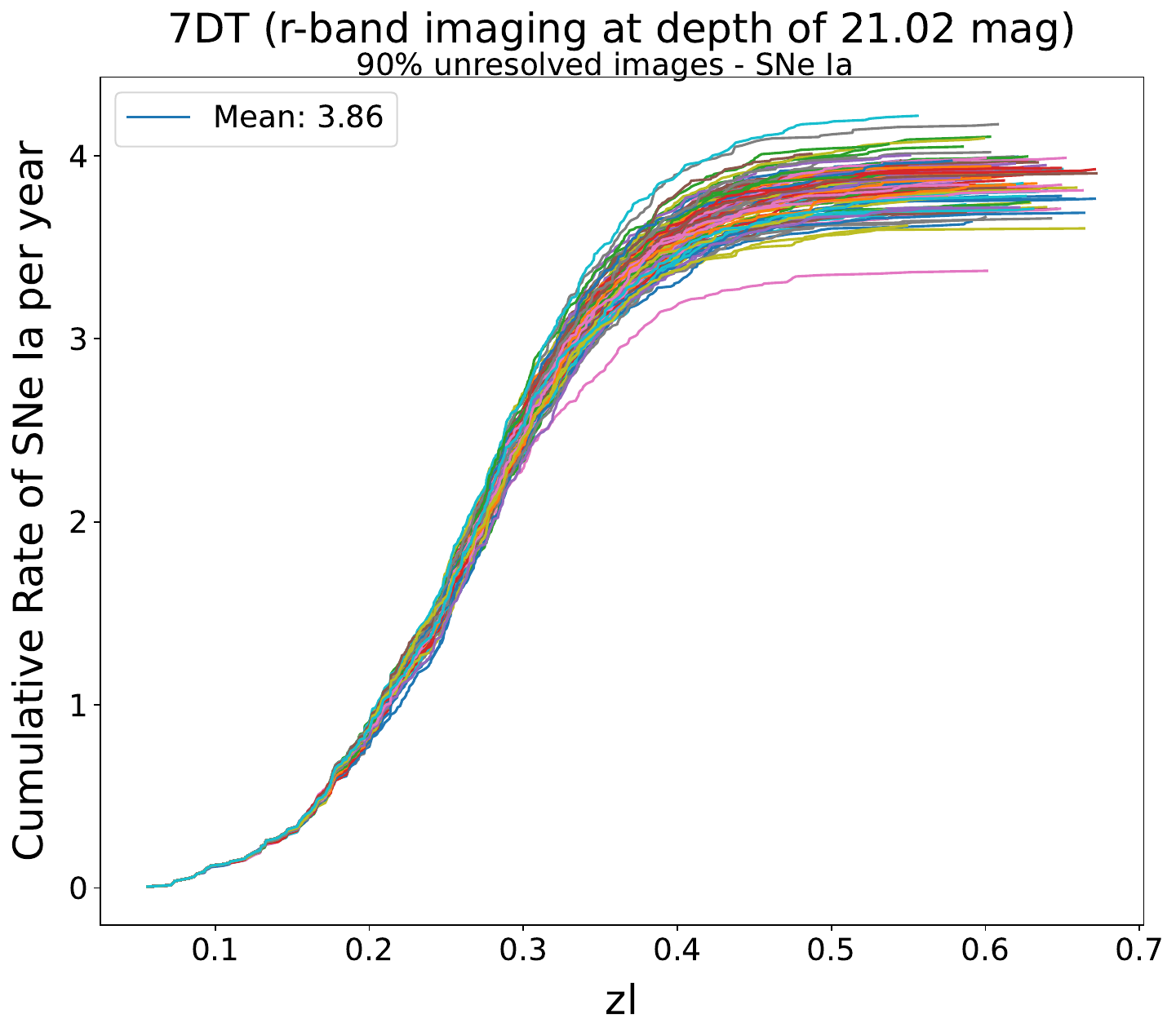}
 \end{minipage}
\begin{minipage}{\textwidth}
\centering
\includegraphics[angle=0,width=0.4\textwidth,clip=]{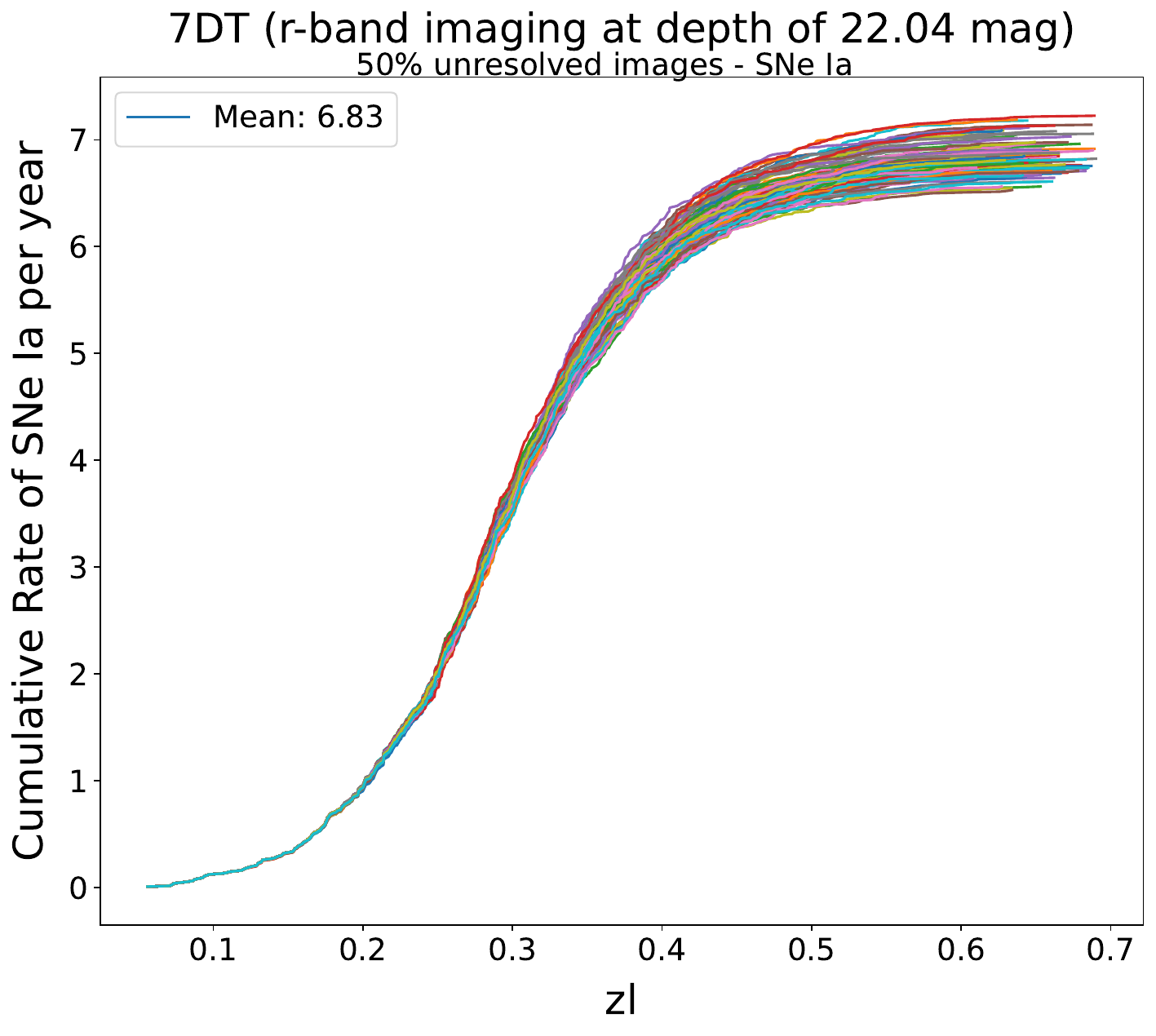}
\includegraphics[angle=0,width=0.4\textwidth,clip=]{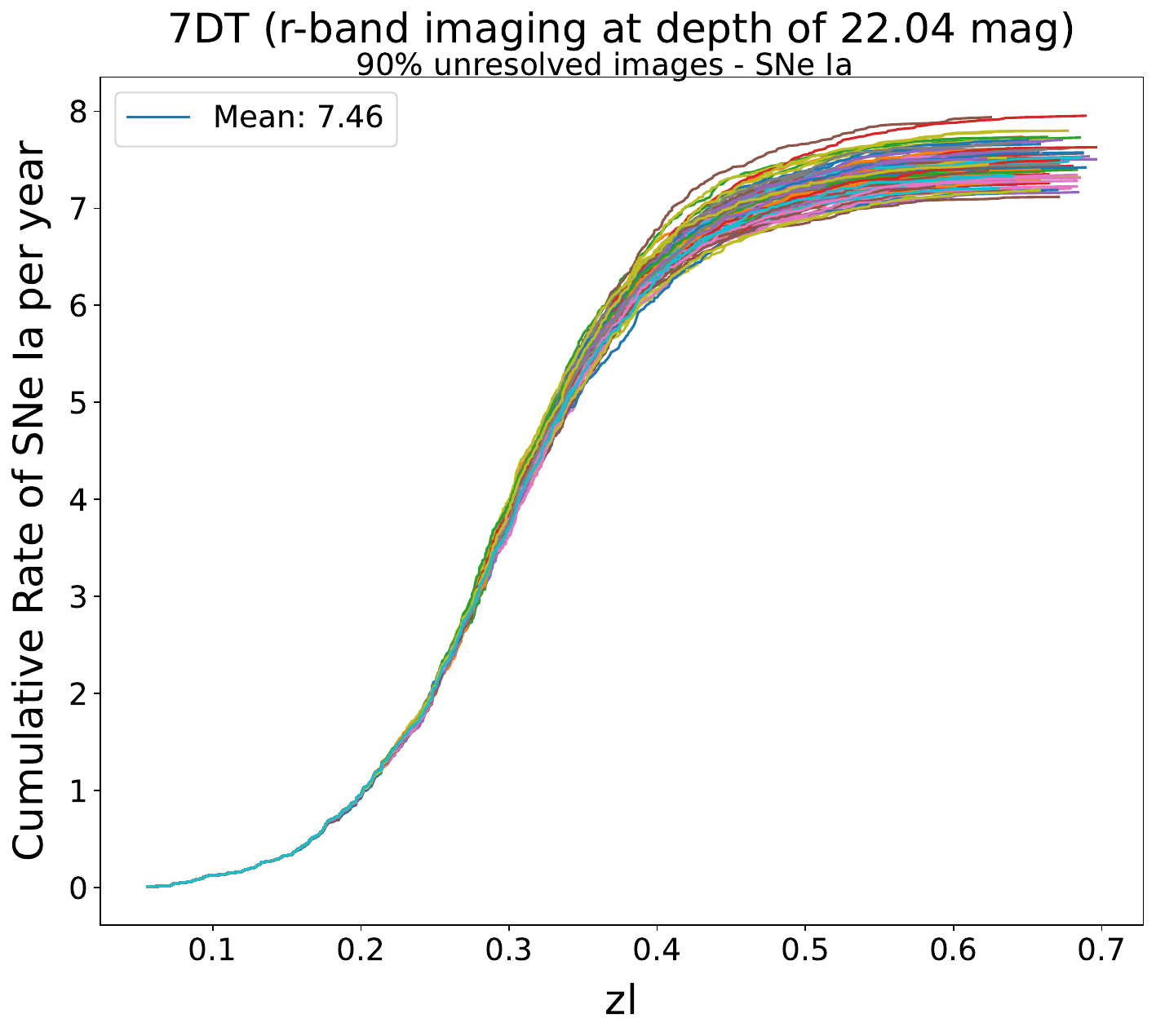}
 
\end{minipage}
\caption{
Upper panels: Control time distribution of glSNe Ia at different observing depths for source redshifts z$_{s}$ $\le$ 0.6 and z$_{s}$ $>$ 0.6, considering (1) 50\% and (2) 90\% of systems are unresolved.  Black vertical lines indicate the median control times.
Lower panel: Annual detection rate of glSNe Ia as a function of lens redshift calculated from 100 simulation realizations (individual realizations shown by colored lines). We present results at different observational depths, accounting for (1) 50\% and (2) 90\% of systems being unresolved.} 
\label{f1}
\end{figure*}

\begin{figure}
\includegraphics[angle=0,width=0.5\textwidth,clip=]{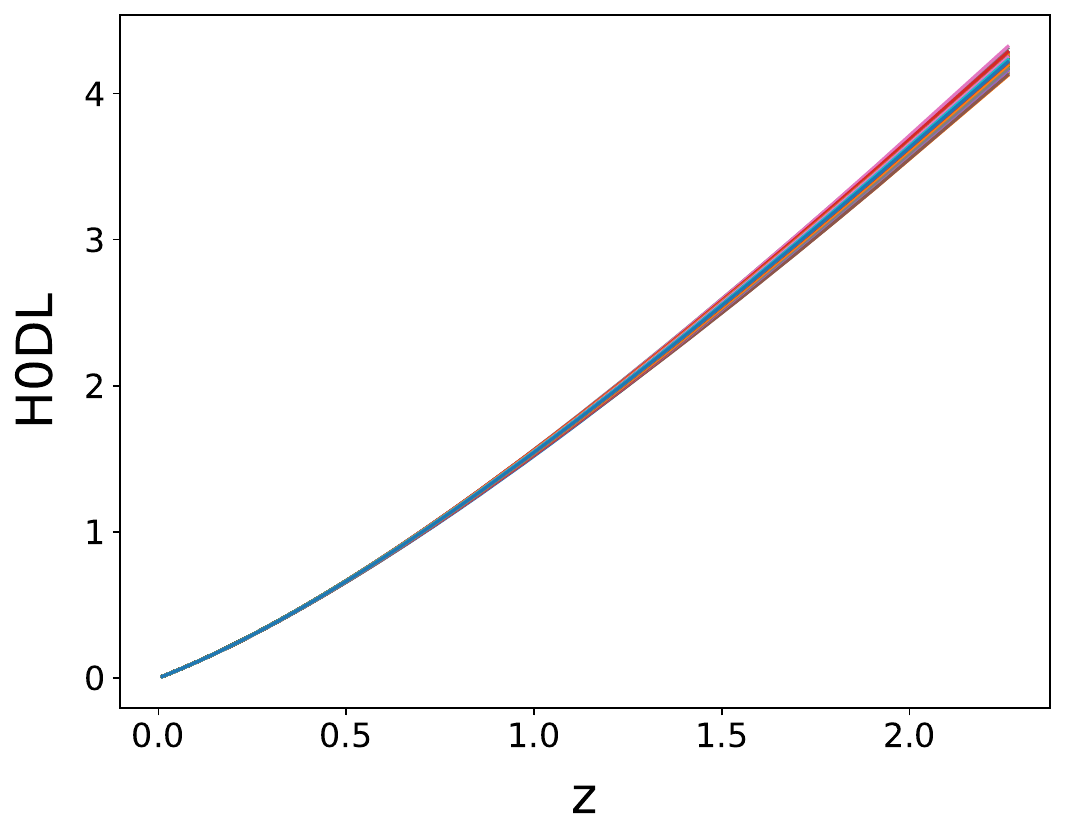}
\caption{ 
Reconstruction of the H0DL(z) function using GP reconstructions from SNe Ia dataset.
}
\label{GP}
\end{figure}

\begin{table}[!ht]
\centering
\caption{Best fit values for $H_{0}$ with 1$\sigma$ uncertainty at $10\%$ precision in time-delay distance measurements of glSNe.}
\label{hubble}
\begin{tabularx}{\columnwidth}{>{\centering\arraybackslash}X|>{\centering\arraybackslash}X|>{\centering\arraybackslash}X}
\hline
\hline
\textbf{Type} & \textbf{$H_{0}$ Best-fit value (1$\sigma$)} & \textbf{Precision}  \\
\hline
7 Ia & $ 69.9 \pm 2.6$  & 3.8\% \\
 
4 IIP  & $ 71.24 \pm 3.6$ & 5.1\% \\
 
2 Ic  & $ 68.05 \pm 4.6$  & 6.8\% \\
 
1 Ib  & $ 69.0 \pm 6.7 $ & 9.8\% \\

7 Ia + 7 CC (Broad-band)  & $ 70.03 \pm 1.9$  & 2.7\% \\
2 Ia (Medium-band)  & $ 71.4 \pm 5.1$ & 7.2\% \\
\hline 
\hline
\end{tabularx}
\end{table}

\end{acknowledgements}

\bibliographystyle{aa}
\bibliography{ref1}

\begin{appendix}

\onecolumn
\section{Control time distribution of glSNe Ia at different 7DT medium-band filters}

\FloatBarrier

%\FloatBarrier  
\begin{figure*}[!h]
\centering
\includegraphics[angle=0,width=0.95\textwidth,clip=]{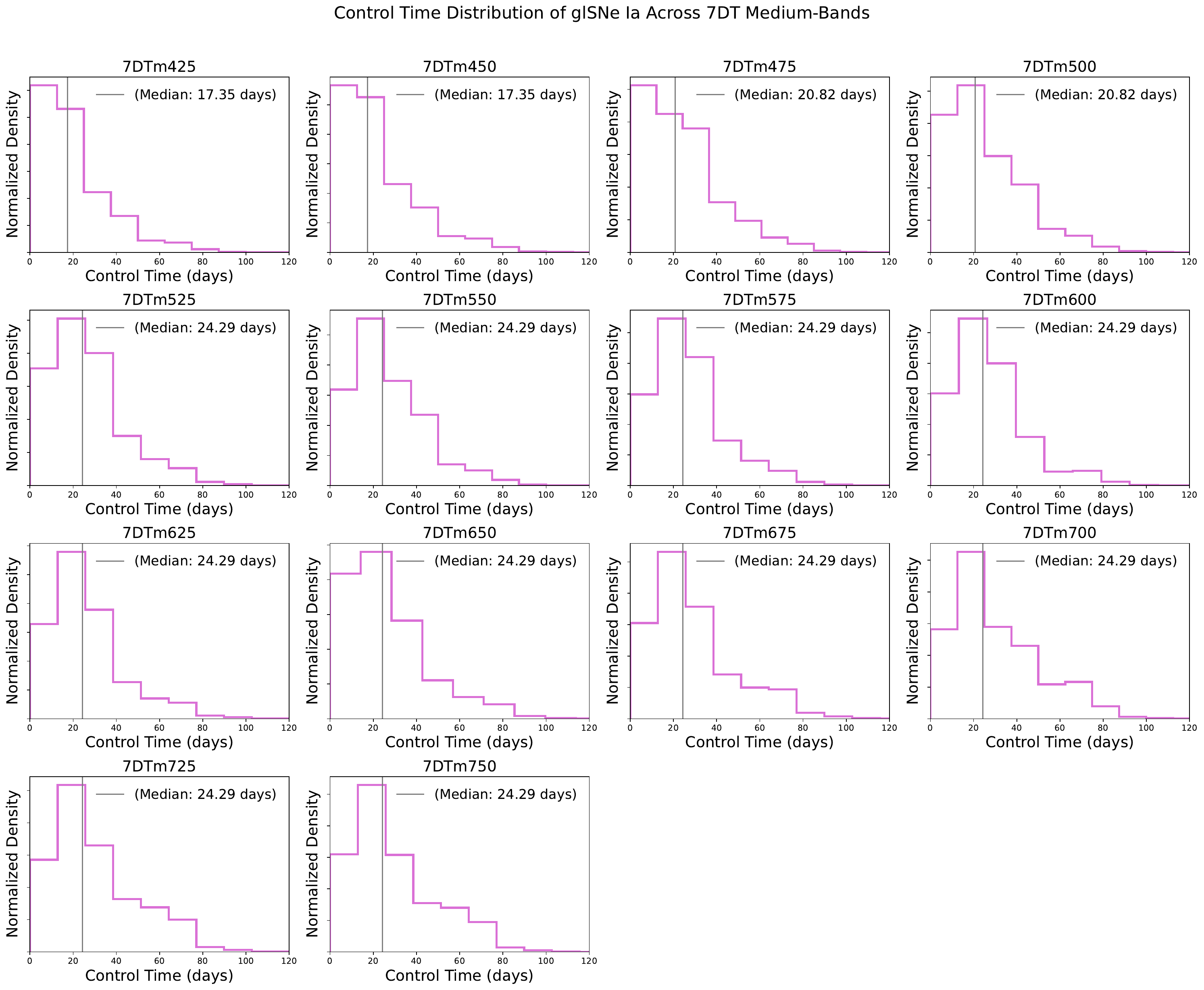}  
\caption{Control time distribution across different 7DT medium-band filters, considering that $90\%$ of systems are unresolved. Median control times are represented by black vertical lines.}
\label{medium_T}
\end{figure*}  

\onecolumn
\section{Control time distribution of glSNe CC at different observing depths}\vspace{-1em}
\FloatBarrier

\begin{figure*}[!h]
\centering
\includegraphics[angle=0,width=1.0\textwidth,clip=]{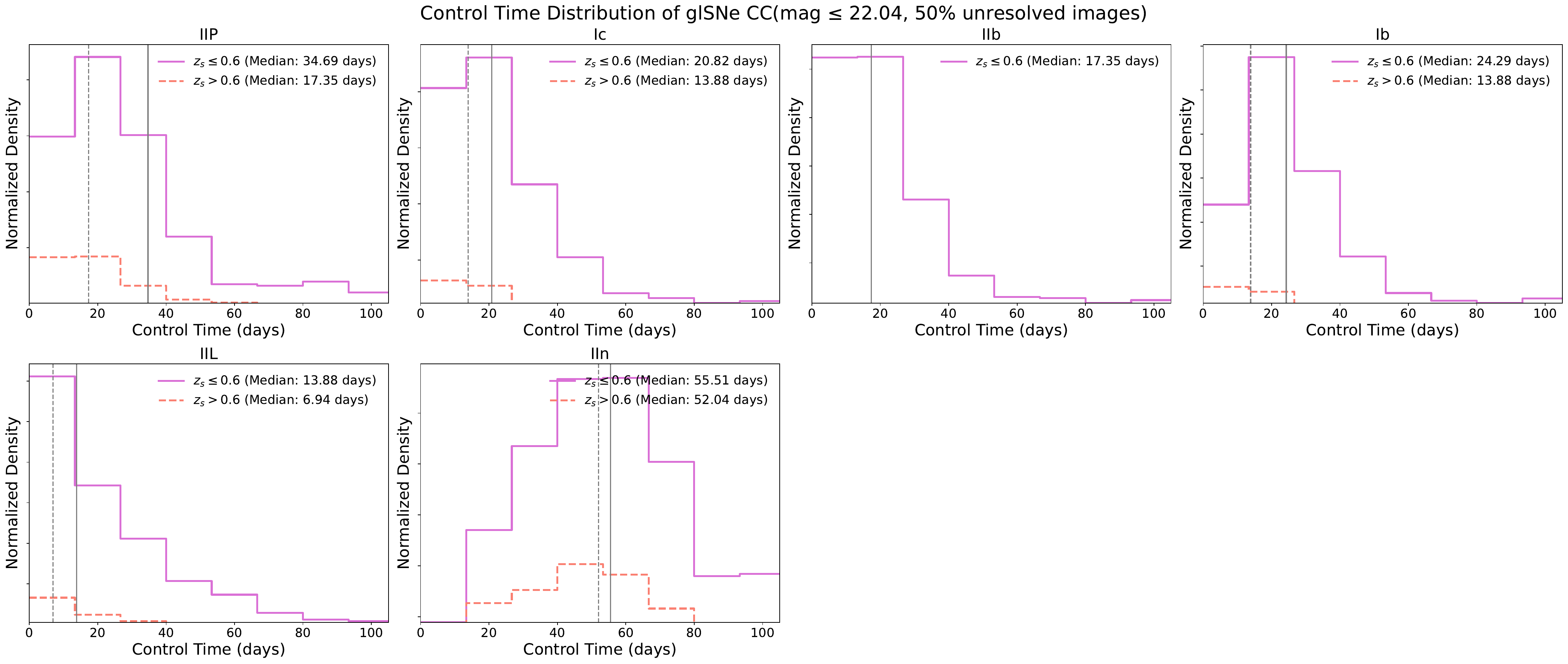}  
\end{figure*} 
\begin{figure*}[!h]
\centering
\includegraphics[angle=0,width=1.0\textwidth,clip=]{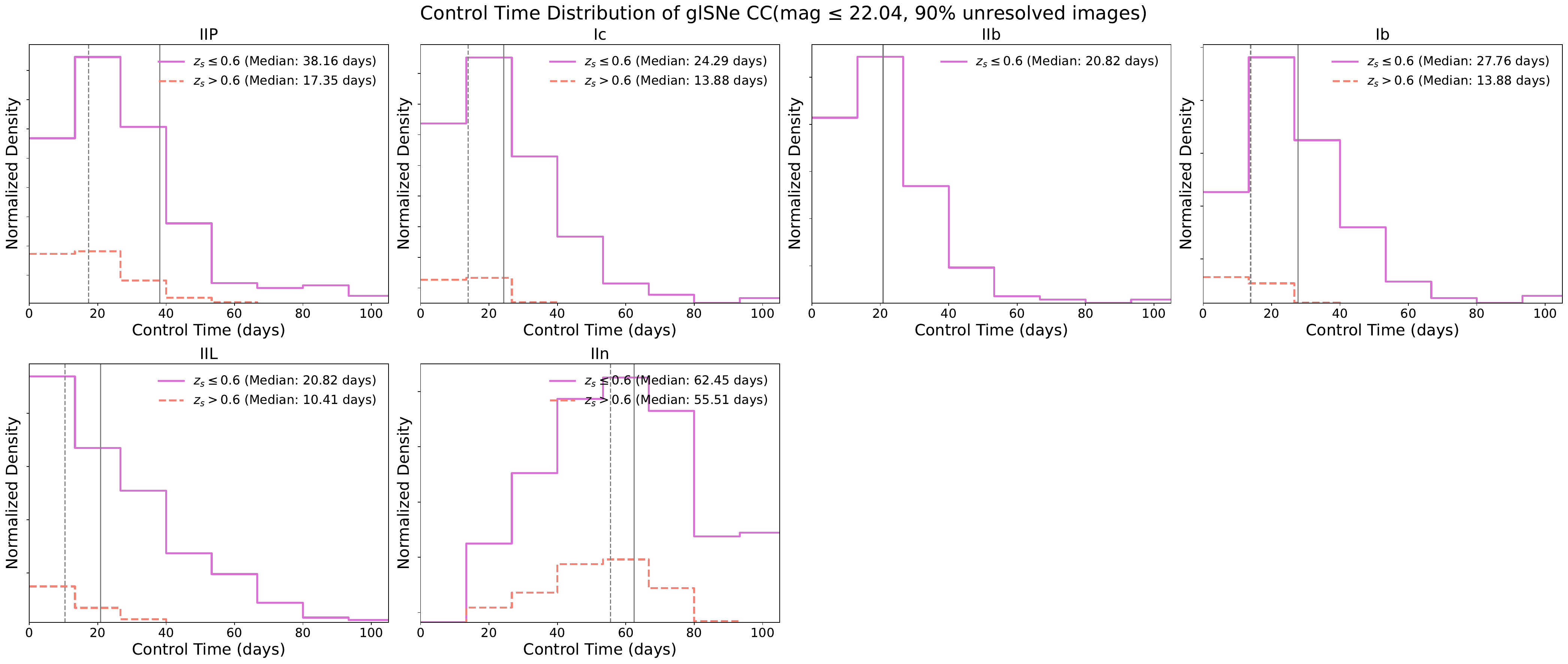}  
\caption{Control time distribution of glSNe CC at observing depths $\leq$ 22.04 for source redshifts z$_{s}$ $\le$ 0.6 and z$_{s}$ $>$ 0.6, considering (1) 50\% and (2) 90\% of systems are unresolved. Median control times are shown with black vertical lines.}
\label{b1}
\end{figure*} 
\begin{figure*}[!h]
\centering
\includegraphics[angle=0,width=1.0\textwidth,clip=]{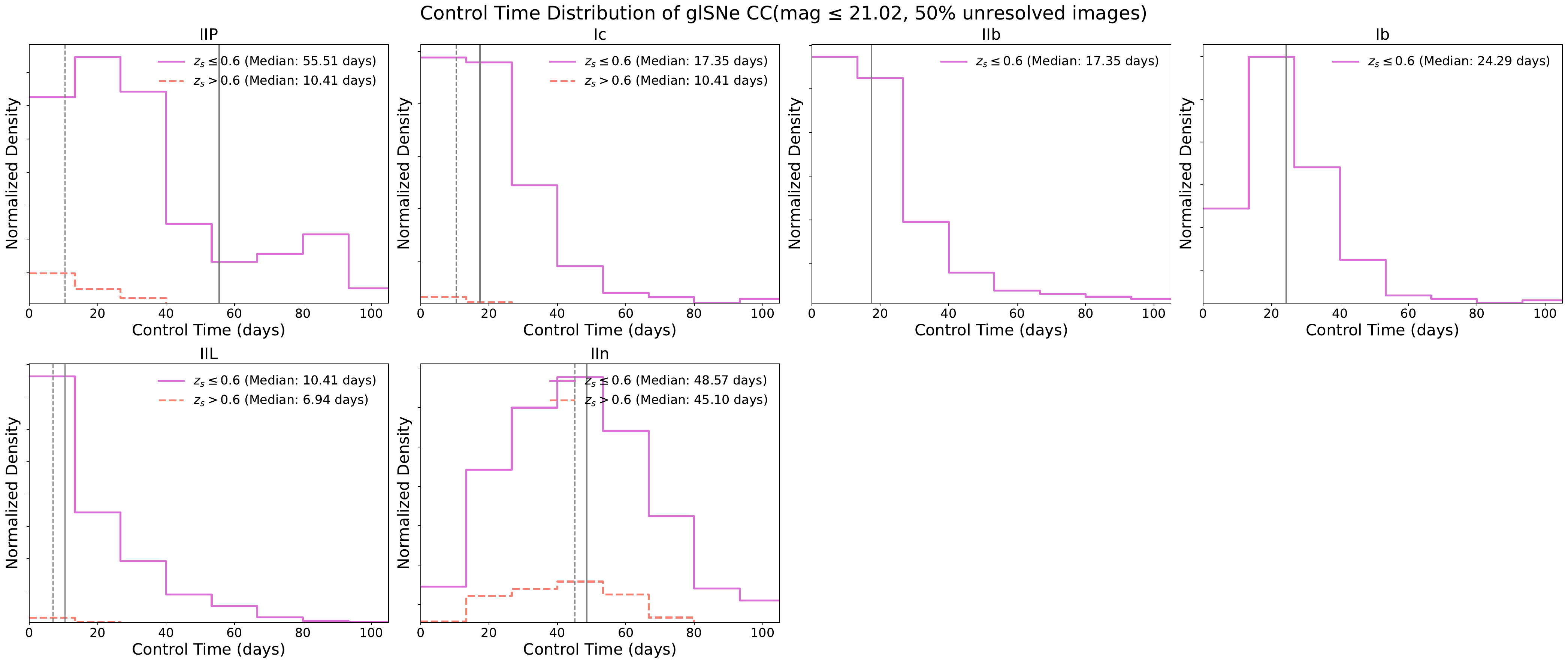}  
\end{figure*} 
\begin{figure*}[!h]
\centering
\includegraphics[angle=0,width=1.0\textwidth,clip=]{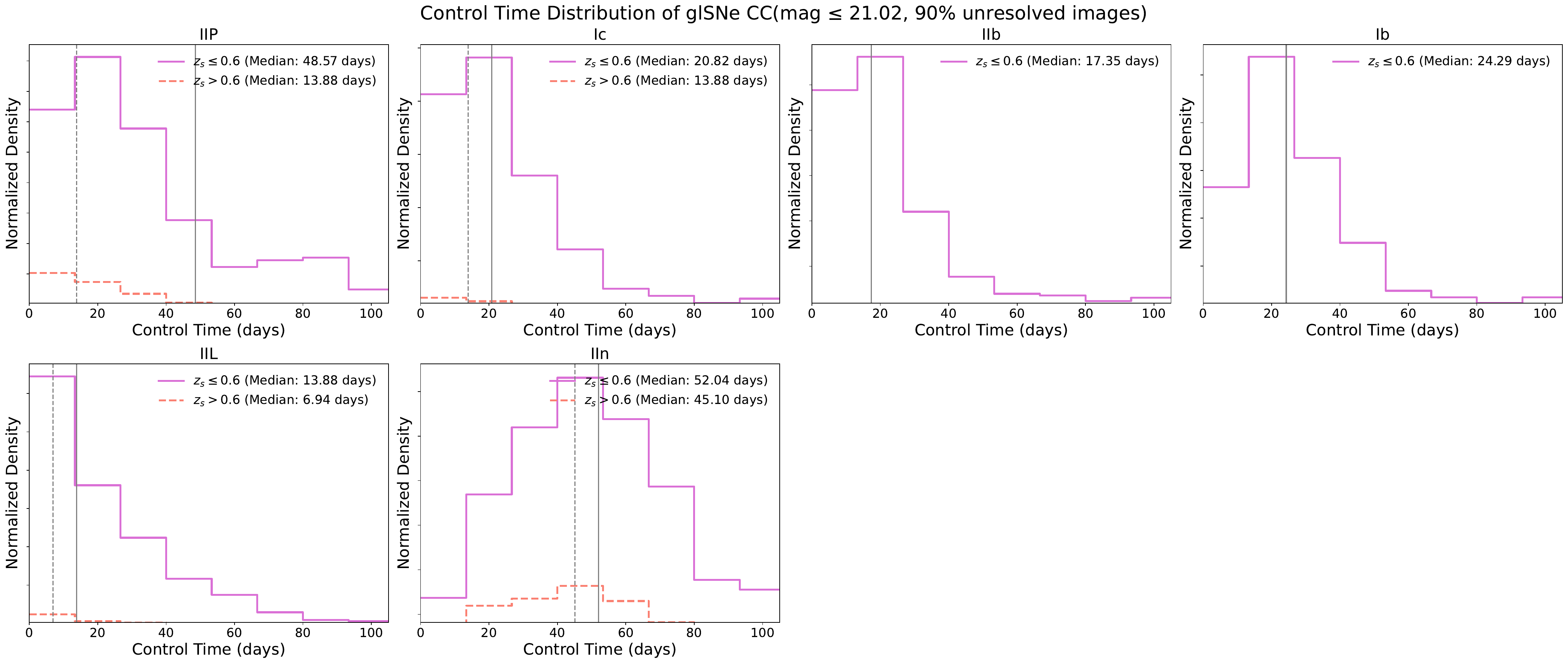}  
\caption{Same as Fig. \ref{b1}, but for observing depths $\leq$ 21.02.}
\label{fcc}
\end{figure*} 

\onecolumn
\section{Annual detection rate of glSNe CC at different observing depths}\vspace{-1em}
\FloatBarrier

\begin{figure*}[!h]
\centering
\includegraphics[angle=0,width=1.0\textwidth,clip=]{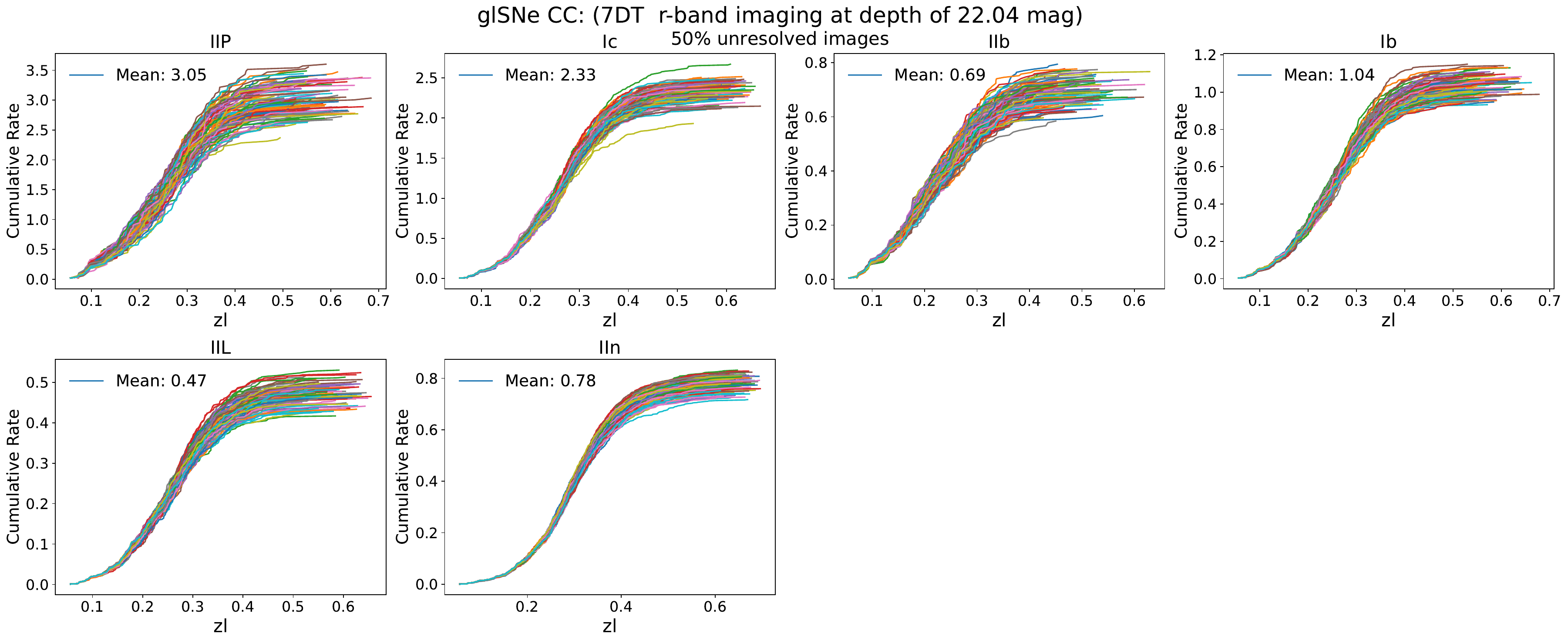}  
\end{figure*} 
\begin{figure*}[!h]
\centering
\includegraphics[angle=0,width=1.0\textwidth,clip=]{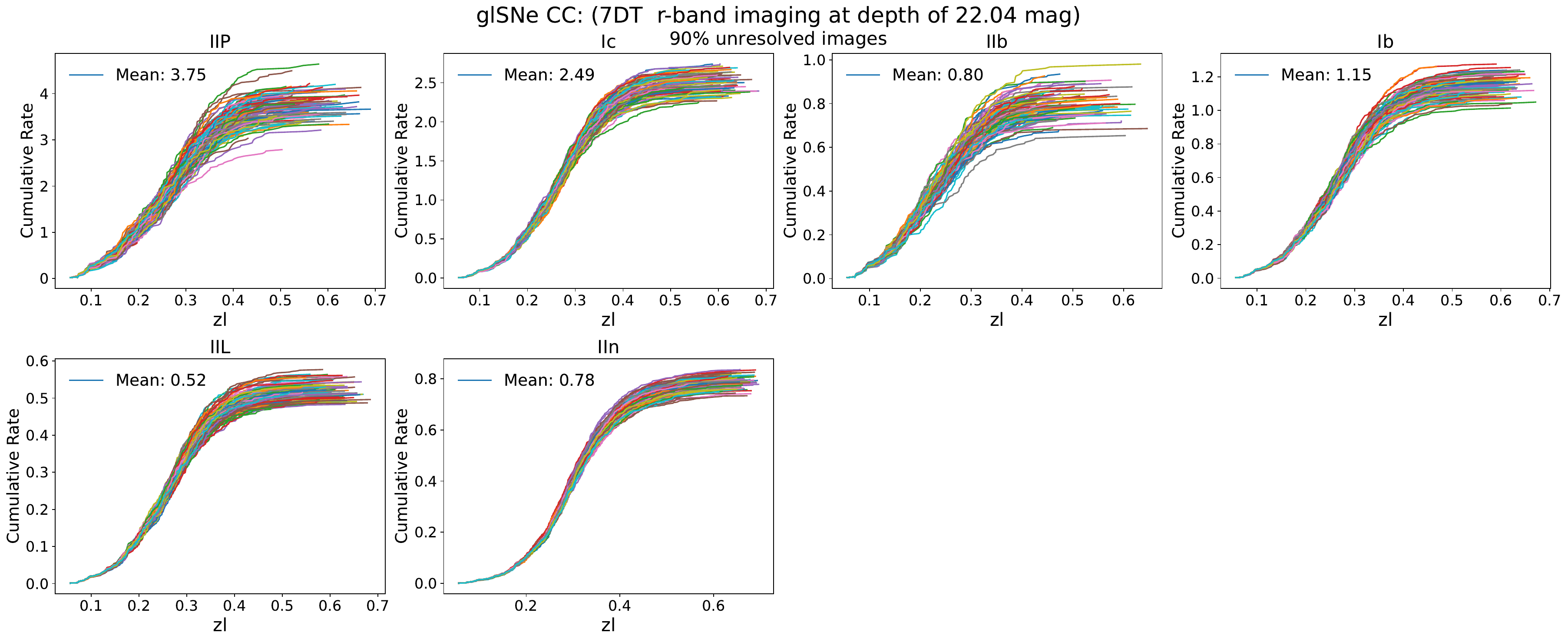}  
\caption{
Annual detection rate of glSNe CC as a function of lens redshift calculated from 100 simulation realizations (individual realizations shown by colored lines). We present results at observational depths $\leq$ 22.04, accounting for (1) 50\% and (2) 90\% of systems being unresolved.} 
\label{c1}
\end{figure*} 

\begin{figure*}[!h]
\centering
\includegraphics[angle=0,width=1.0\textwidth,clip=]{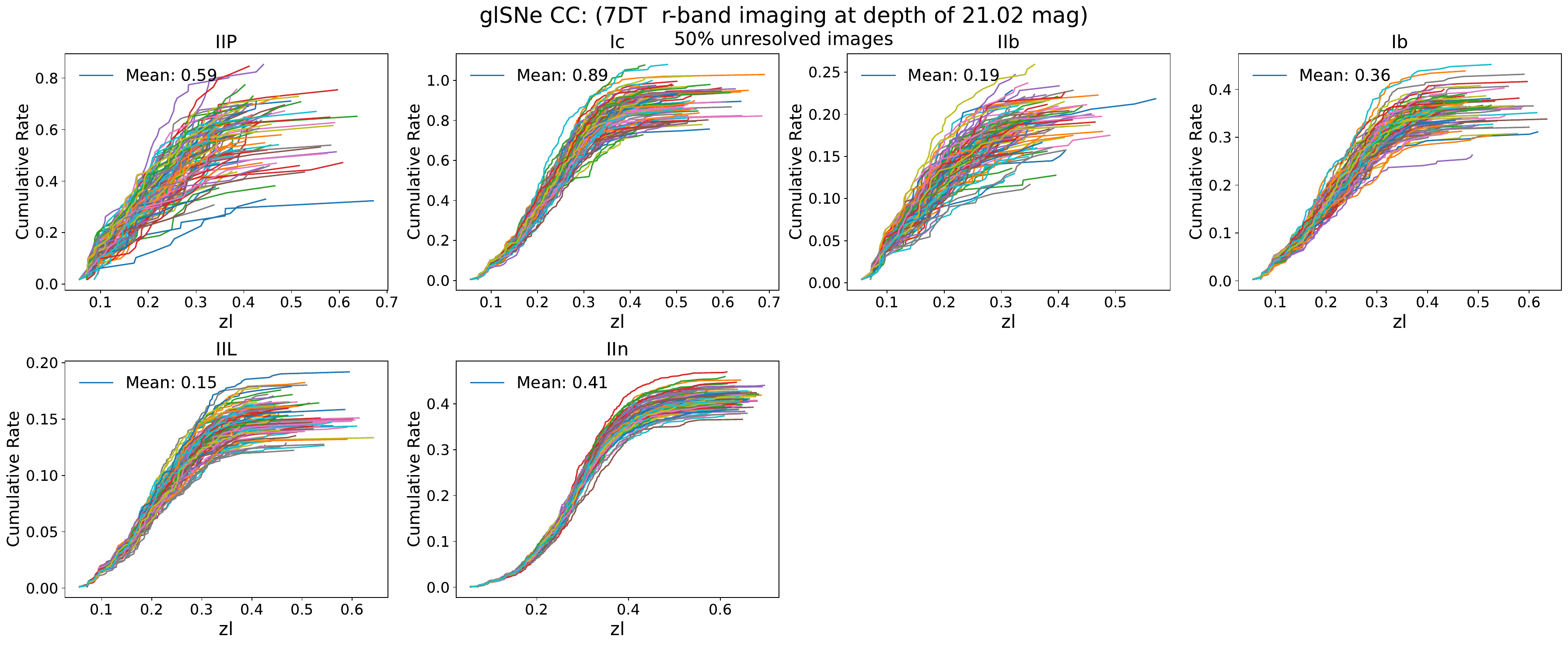}  
\end{figure*} 
\begin{figure*}[!h]
\centering
\includegraphics[angle=0,width=1.0\textwidth,clip=]{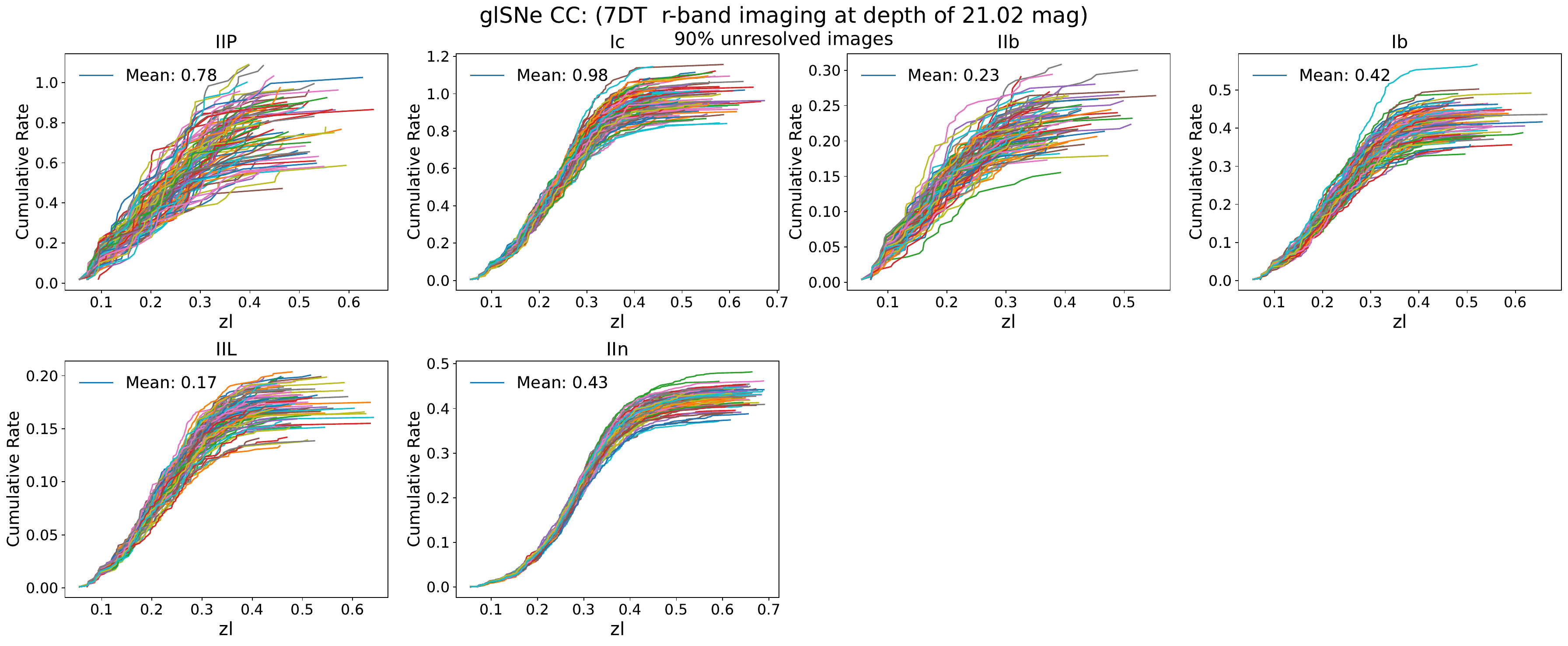}  
\caption{Same as Fig. \ref{c1}, but for observing depths $\leq$ 21.02.}
\label{fcc-rate}
\end{figure*} 
\end{appendix} 

\end{document}